\documentclass[prx,showkeys,showpacs,nofootinbib,,twocolumn,unsortedaddress]{revtex4-2}
\usepackage{graphicx} 
\usepackage{dcolumn} 
\usepackage{amssymb}
\usepackage{amsmath}
\usepackage{physics}
\usepackage{natbib}
\usepackage{xcolor}
\usepackage{soul}
\usepackage{siunitx}
\usepackage{float}
\usepackage{esdiff}
\usepackage{bbold}
\usepackage{mathtools}
\usepackage{todonotes}
\usepackage{hyperref}
\usepackage{lineno}

\colorlet{RED}{red}

\begin{document}

\title{Electrostatic splitting of an Edge Magnetoplasmon Resonator}

\author{Slo\"an Kouamé$^{1\dagger}$}
\author{Elric Frigerio$^{1\dagger}$}
\author{Giacomo Rebora$^{2}$}
\author{Suvankar Purkait$^{1}$}
\author{Yong Jin$^{3}$}
\author{Ulf Gennser$^{3}$}
\author{Antonella Cavanna$^{3}$}
\author{Jean-Marc Berroir$^{1}$}
\author{Emmanuel Baudin$^{1}$}
\author{Pascal Degiovanni$^{2}$}
\author{François D. Parmentier$^{1}$}
\author{Gwendal F\`eve$^{1}$}
\author{Gerbold C. M\'enard$^{1}$}
\email{email: gerbold.menard@phys.ens.psl.eu}
\affiliation{$^{1}$ Laboratoire de Physique de l’Ecole normale sup\'erieure, ENS, Universit\'e PSL, CNRS, Sorbonne Universit\'e, Universit\'e Paris Cit\'e, F-75005 Paris, France}
\affiliation{$^{2}$ Univ Lyon, ENS de Lyon, Université Claude Bernard Lyon 1, CNRS, Laboratoire de Physique, F-69342 Lyon, France}
\affiliation{$^{3}$ Centre de Nanosciences et de Nanotechnologies (C2N), CNRS, Universit\'e Paris-Saclay, 91120 Palaiseau, France}

\begin{abstract}

Edge-magnetoplasmon resonators have been proposed as a powerful tool to detect anyons by introducing a quantum point contact into an isolated quantum Hall system probed via radiofrequency radiation. In this paper, we study the effect of a quantum point contact embedded within an edge-magnetoplasmon resonator and how its polarization influences the propagating magnetoplasmonic mode. Combining dc and rf measurements, we unambiguously evidence the signature of both integer ($\nu = 1$ and $2$) and fractional quantum Hall states ($\nu = 4/3$ and $2/3$) within the radiofrequency transmission signal. Using electrostatic gating, we determine the physical parameters characterizing the electrostatic edge of an AlGaAs/GaAs based two-dimensional electron gas. We extract the dependence of the cavity perimeter with the gate voltage of the quantum point contact and fully characterize the path followed by edge magnetoplasmons in this system. Finally, we provide a geometric model in good agreement with experimental results.

\end{abstract}


\date{\today}

\maketitle
\def\thefootnote{$\dagger$}\footnotetext{These authors contributed equally to this work}\def\thefootnote{\arabic{footnote}}

\section{Introduction}

Edge-magneto plasmons (EMPs) are plasmonic excitations of quantum Hall systems built using the one-dimensional edge modes that carry the Hall current \cite{andrei1988low, volkov1988edge, safi1995, von1998, safi1999, senechal1999}. An isolated quantum Hall droplet behaves as a microwave resonator where the resonance condition is obtained through quantization of the wave vector \cite{talyanskii1990edge, talyanskii1992spectroscopy, balaban1997observation, hashisaka2013, kumada2014resonant}.

EMP resonators have been used to detect the state of double quantum dots through the modulation of the radiofrequency (rf) amplitude \cite{lin2024resonant, lin2026dispersive}. They have also been used as rf circulators \cite{bosco2019, mahoney2017chip, martinez2024edge, tarascio2026compact} able to function at non-zero magnetic fields. The interest of such devices lies in the high-impedance of the system making it relevant for coupling to other high-impedance devices with lower losses \cite{bosco2017self, fijalkowski2024balanced}. In particular this presents an interest for the high-frequency study of quantum Hall systems.

In a previous work \cite{frigerio2024} we studied the electrostatic control of gated EMP cavities and have shown that the combination of the magnetic field and electrostatic fields allows to reliably control the resonance frequency of these cavities. The electric field controls both the geometry of the cavity and the electronic density of the system, while the magnetic field tunes the velocity of edge states.

Our goal in this paper is to study the effect of a quantum point contact (QPC) integrated in such structure. This study is motivated by the theoretical prediction of the realization of an EMP interferometer that would potentially allow for the detection of anyonic excitations in the fractional quantum Hall regime \cite{cano2013microwave, microwave2025marseille}. When applying a negative electrostatic voltage, the QPC splits the quantum Hall droplet into two lobes and should therefore lead to interference effects by generating tunneling between counter-propagating edge states close to the pinch-off point.

In this paper, we first use electrostatic gating in order to study electrostatic parameters of a two-dimensional electron gas (2DEG). More precisely, we are interested in two main physical parameters: the dielectric permittivity of the top AlGaAs layer and the electronic depletion length at the edge of the droplet. By first analyzing the dc transport in the system we can reliably identify rf resonances associated to the fractional quantum Hall regime. Accessing both the density and the field dependence in rf finally allows reconstructing the path followed by EMPs within the cavities and quantitatively extracting the corresponding perimeter as the QPC gets fully pinched.



\section{Sample description}
\begin{figure*}[t]
    \begin{center}
    \includegraphics[width=0.9\linewidth]{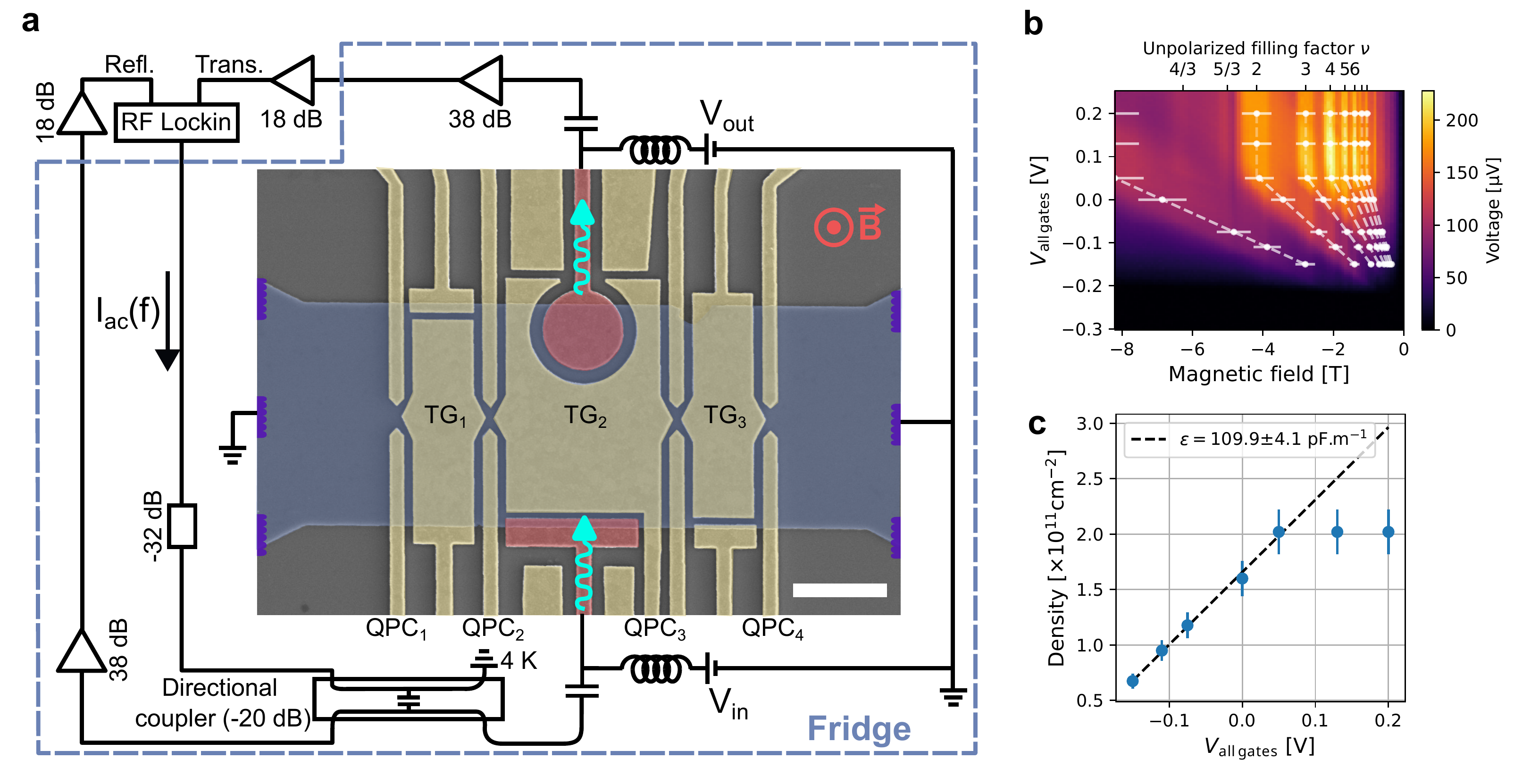}
    \caption{\textbf{Sample and measurement apparatus:} The sample is constituted of a 2DEG covered by a set of electrostatic (yellow) and rf (red) gates. Ohmic contacts are also present at the edge of the sample and are simply schematically represented here (purple). The input rf signal is sent from a radiofrequency lock-in amplifier and passes through distributed attenuators (-32 dB total attenuation) and a directional coupler (-20 dB attenuation). The signal then has two possible paths, either go through the sample where it is collected by a first set of amplifiers, or is reflected and collected by the second set of amplifiers and sent to another input of the lock-in amplifier. The applied magnetic field is perpendicular to the surface of the sample. The white scale bar represents a length of \SI{2}{\micro\meter}. \textbf{(b)} DC voltage transmitted through the device as a function of the homogeneous voltage applied on all gates and the external magnetic field. The white crosses correspond to the center of Hall plateaus with the corresponding error bars estimated as 10\% of the magnetic field. \textbf{(c)} Corresponding density extracted from the position of plateaus as a function of the gate voltage. The dashed line represents the linear fit of the data from which we extract the dielectric permittivity $\varepsilon = 109.9\pm4.1$~\SI{}{\pico\farad\per\meter}. The error bars are propagated from the error on the position of plateaus shown on panel b. All data was acquired at the base temperature of the fridge of \SI{30}{\milli\kelvin}.}
    \label{fig1_sample}
    \end{center}
\end{figure*}

The experiments described below were performed on samples based on AlGaAs/GaAs two-dimensional electron gases. Those samples were etched chemically and covered with electrostatic gates as shown in figure \ref{fig1_sample}. The injection and detection of rf signal is performed via two gates shown in red in figure \ref{fig1_sample}.a. These specific gates are addressable both via dc and rf owing to the presence of bias tees. Similarly to the work presented in \cite{frigerio2024}, we use the electrostatic gates to define different configurations of our resonator. Four quantum point contacts (QPCs) allow us to define six different cavities, out of which only four different measurable cavities that contain the input and output gates.  Finally, we deposit and anneal ohmic contacts in order to characterize the dc properties of our system.

On the rf side, the output gate is simply connected to a cryogenic amplifier at low temperature before reaching a room temperature amplifier and finally to the radiofrequency lock-in amplifier input port. The input gate is connected to the output port of the lock-in via a directional coupler which allows to also measure the reflected signal of the cavity. The total attenuation of the input line, accounting for the \SI{-20}{\decibel} of the directional coupler, is of \SI{-52}{\decibel}. The top output port of the directional coupler connects to the ground at the \SI{4}{\kelvin} stage of the fridge in order to dissipate the incoming heat due to the \SI{20}{\decibel} attenuation in the directional coupler far from the mixing chamber at a stage with higher cooling power.

\section{Experimental results}
    \subsection{DC characterization}

We start our study by focusing on the dc properties of the device, using the ohmic contacts shown in purple on figure \ref{fig1_sample}.a. We apply a dc voltage from the bottom right ohmic contact to the bottom left one. The transmitted signal is then measured as a function of the perpendicular magnetic field and the voltage of all electrostatic gates. In this configuration, all top gates ($\mathrm{TG}_{1,2,3}$) and QPCs ($\mathrm{QPC}_{1,2,3,4}$) are polarized at the same voltage $V_\mathrm{all~gates}$. This is equivalent to a single wide top gate that can tune the electronic density of the 2DEG. The result of such measurement is given in figure \ref{fig1_sample}.b.


On this figure, three regimes can be observed. First for gate voltages $V_\mathrm{all~gates}>0.1$\SI{}{\volt}, we observe vertical lines corresponding to the quantum Hall plateaus. In this region, the gates are fully open and pushing them toward larger voltages does not alter the transmission of the device as the 2DEG below the gates reach a larger electronic density than the ohmic access. Because the resulting additional channels do not contribute to the electric transport through the cavity the quantum Hall plateaus do not show any more dependence with the gate voltage. From this point, we can infer the electronic density (see figure \ref{fig1_sample}.c) of the ungated region of the device to be $n_\mathrm{ungated} = 2.02\times 10^{11}$\SI{}{\per\centi\meter\squared} and identify each feature to a specific value of the filling factor. In particular, we observe the signature of two fractional states $\nu = 4/3$ and $\nu = 5/3$ in this region of the data.

\begin{figure}[t]
    \begin{center}
    \includegraphics[width=0.8\linewidth]{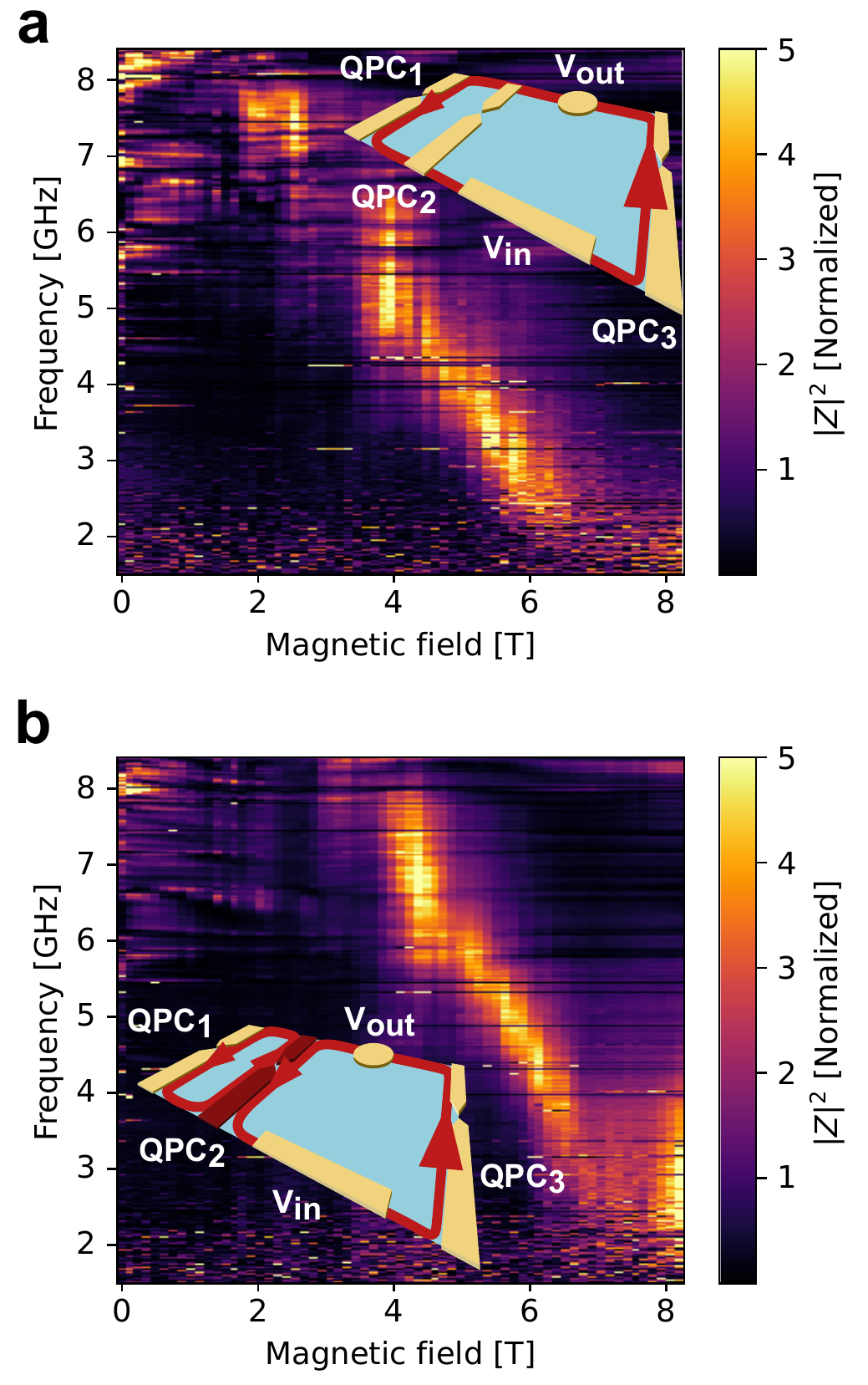}
    \caption{\textbf{RF characterization of the device:} \textbf{(a)} Transmission map of the cavity in the "middle left" configuration. In this situation, $\mathrm{QPC}_{1}$ and $\mathrm{QPC}_3$ are pinched-off. \textbf{(b)} Same in the "small" configuration. In this case, $\mathrm{QPC}_2$ is also pinched-off, therefore isolating the left part of the device from the input and output gates. The insets show the cavity configurations schematically where the top gates have been removed for clarity.}
    \label{rf characterization main}
    \end{center}
\end{figure}

At voltages below \SI{-0.2}{\volt}, the combined actions of the top gates fully pinches the device and therefore the transmission falls to zero as the 2DEG is fully depleted and thus isolates the source from the drain. In between these two extreme cases, we observe a linear shift of the plateaus that correspond to a gradual reduction of the 2DEG density in the gated regions. From this shift we can fit the position of the plateaus as a function of the gate voltage and plot figure \ref{fig1_sample}.c where the density is represented as a function of $V_\mathrm{all~gates}$. 

In the approximation of a planar capacitor, we expect the density to follow the relation:
\begin{equation}
    n_B = \frac{(V_G-V_0)C_G}{eS},
\end{equation}
where $C_G$ is the capacitive coupling between the gate and the 2DEG, $S$ the total surface of the gate and $e$ the electron charge. $V_0$ is the offset voltage. The capacitance follows in this model the relation $C_G = \varepsilon S/d$ where $d=$~\SI{105}{\nano\meter} is the thickness of the AlGaAs separating the gate from the 2DEG. From a linear fit of the data in figure \ref{fig1_sample}.c we can thus extract the dielectric permittivity of the AlGaAs layer $\varepsilon = 109.9\pm 4.1$\SI{}{\pico\farad\per\meter}$=12.4\times\varepsilon_0$. It corresponds to the value reported in the literature \cite{strzalkowski1976, moore1996infrared, krupka2008} and the one used in our previous work for simulation purposes \cite{frigerio2024}.

\subsection{RF characterization}

Our system can accommodate four distinct cavity configurations by polarizing different QPCs. Closing QPCs 1 and 4 defines the "large" configuration, closing QPCs 1 and 3 defines the "middle left" configuration, closing QPCs 2 and 4 defines the "middle right" configuration and closing QPCs 2 and 3 defines the "small" configuration. We do not use the "large" and "middle-right" configurations in the context of this paper.

We present in figures \ref{rf characterization main}.a and b the rf transmission maps as a function of magnetic field and frequency in the "middle left" and "small" cavity configurations respectively (see inset schematics). The quantity $|Z|^2$ plotted on these maps is the square normalized transmission amplitude, where the input gate is the red bottom gate shown in figure \ref{fig1_sample}.a and the output gate is the round gate shown in red in the same figure. This normalization procedure is described in detail in \cite{frigerio2024}. Both maps show clear resonance signals and the shift of the resonance frequency is consistent with the different dimensions of the cavities (\SI{22}{\micro\meter} and \SI{18}{\micro\meter}). These maps were acquired for a top gate voltage (that includes TG$_1$, TG$_2$ and QPC$_2$) of $V_\mathrm{all~gates} = -75$~\SI{}{\milli\volt}. On these maps, we observe a variation of the signal amplitude as a function of the magnetic field. In particular, two main bright spots can be observed at $B\simeq$~\SI{5.5}{\tesla} and $B\simeq$~\SI{4}{\tesla} on both maps and, as we will show in the next section, can be respectively attributed to quantum Hall plateaus $\nu = 1$ and $\nu = 4/3$ (see also supplementary section F).
    
\begin{figure*}[t]
    \centering
    \includegraphics[width=0.8\linewidth]{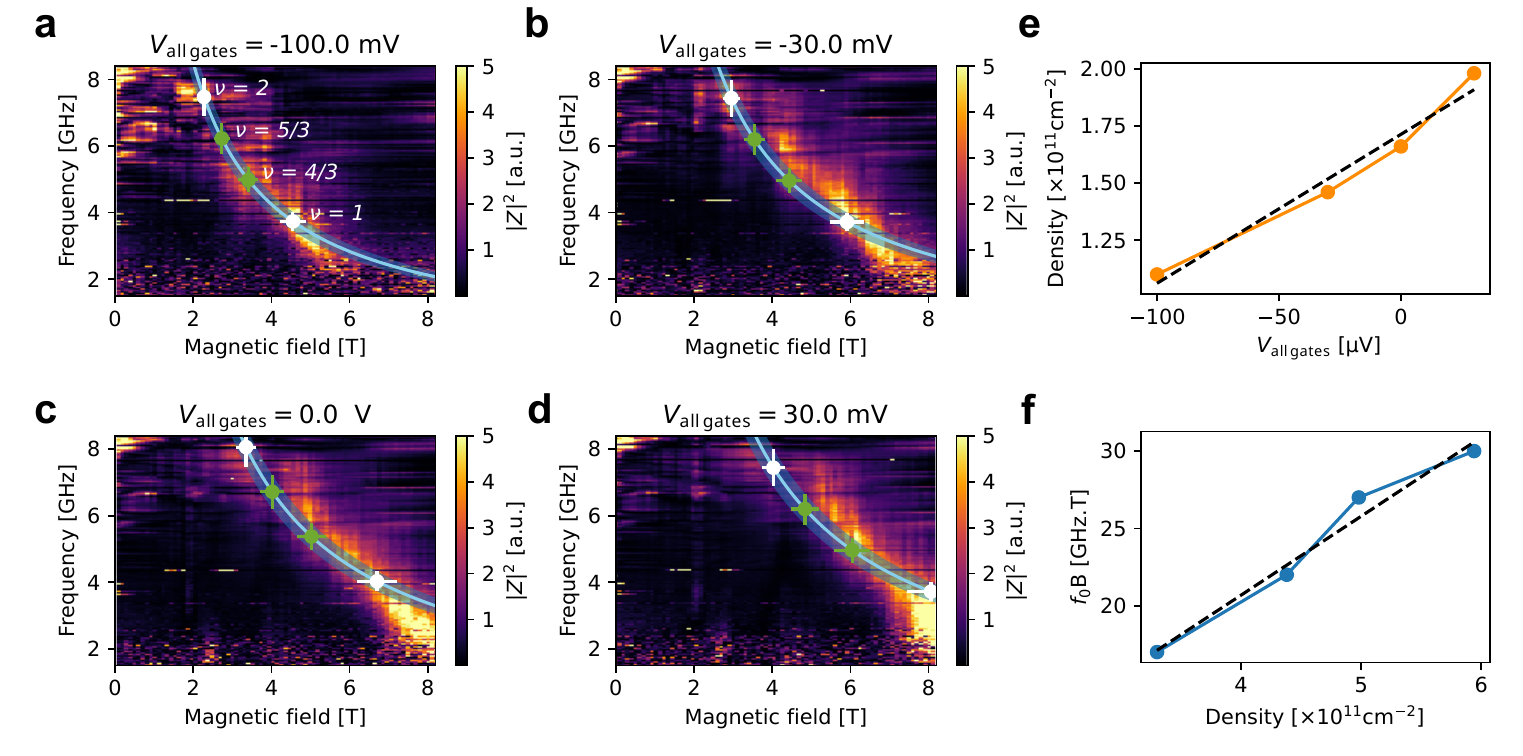}
    \caption{\textbf{Estimation of gate capacitance:} \textbf{(a)-(d)} Frequency vs. Magnetic field maps of the transmission through the cavity in the "middle left" configuration (see inset of figure \ref{rf characterization main}.b) for various polarization $V_\mathrm{all~gates}$ of gates TG$_1$, TG$_2$ and QPC$_2$ (\SI{-100}{\milli\volt}, \SI{-30}{\milli\volt}, \SI{0}{\milli\volt} and \SI{30}{\milli\volt} respectively). The white dots correspond to the position of the integer plateaus and the green ones to the fractional ones. The full lines correspond to the fit of $f_0B$ that follows the $1/B$ trend of the resonance frequency. \textbf{(e)} Extracted value of the electronic density (based on the position of the intensity maxima). The dashed line indicates the linear fit of the density dependence on the gate voltage from which we estimate the gate capacitance $C_\mathrm{G} = neS/V_\mathrm{G}$, dielectric permittivity $\varepsilon_r= C_\mathrm{G}d/S$. \textbf{(f)} Extracted value of $f_0B$ (based on the global shape of the resonance) as a function of the density. The dashed line is a linear fit of the two parameters from which we extract the depletion length $a$ (see text).}
    \label{fig_gate_capacitance}
\end{figure*}

We now set our cavity in the "middle left" configuration by closing QPCs 1 and 3. We then measure four frequency vs. magnetic field maps at four different values of $V_\mathrm{all~gates}$. This time the gates that are polarized are all the gates in the cavity exception made of the input gate, output gate and cavity-defining QPCs. This transmission measurement then leads to the data shown in figures \ref{fig_gate_capacitance}. Our first observation is that of the shift of the resonance toward higher fields and frequency as the gate voltage is increased. This is easily understood as the velocity of the EMPs is directly proportional to the electronic density in the case of gated sample \cite{johnson2003}.

As for figures \ref{rf characterization main}, on each one of these maps, we can observe variations of the amplitude of the resonance signal. We attribute these variations to the presence of the quantum Hall plateaus. As the field is varied from one Hall plateau to another one, the edge modes are not well-defined anymore and thus neither is the resonance \cite{mahoney2017chip, frigerio2024}. We match the brightest areas on the experimental data to quantum Hall plateaus (see supplementary section D and F) which allows us to identify resonances associated to integer Hall states ($\nu = 1$ and 2). We repeat our previous analysis of density dependence with the gate voltage. We plot on figures \ref{fig_gate_capacitance}.a-d the position of these plateaus using dots. The white dots correspond to the integer plateaus and the green ones to fractional plateaus. We then extract the associated electronic density which results in the points shown on figure \ref{fig_gate_capacitance}.e where the dashed line is the linear fit from which we extract the value of the permittivity, using only our rf data. Repeating our previous analysis we obtain $\varepsilon = (12.4\pm 1.7)\times \varepsilon_0$ in perfect agreement with the dc measurement.

This agreement is a confirmation that our identification of quantum Hall plateaus in rf is accurate and that we clearly identify signatures of both integer and fractional plateaus. However, some plateaus exhibit a stronger contrast. In particular, the most visible ones are the $\nu = 1$ and $\nu = 4/3$ plateaus.

It was shown in \cite{frigerio2024} that the resonance appears in our system when $f_0=v/L$ where $v$ is the velocity of EMPs and $L$ is the perimeter of the cavity from which is subtracted the width of the input and output gate. Mathematically, this is easily seen from the expression of the transmission coefficient (see supplementary section B)
\begin{equation}
    S_{ca} = \frac{t_a' t_ce^{iX_b}}{1-r_a'r_c'e^{2iX_b}},
\end{equation}
where $X_b = k(\omega)L_b$ with $L_b$ the length of the free propagation between input and output gates (assumed symmetric). The resonance is obtained when the denominator of this expression is minimized (see supplementary section C) which corresponds to $2iX_b=2n\pi$. This resonance condition thus translates into
\begin{equation}
    f=\frac{v}{2L_b}=\frac{v}{P-L_a-L_c}
    \label{resonance_equation}
\end{equation}
where $L_a = $~\SI{2.8}{\micro\meter} is the width of the input gate, $L_c=$~\SI{1.2}{\micro\meter} that of the output gate and $P$ is the perimeter of the cavity.

\begin{figure*}
    \centering
    \includegraphics[width=0.8\linewidth]{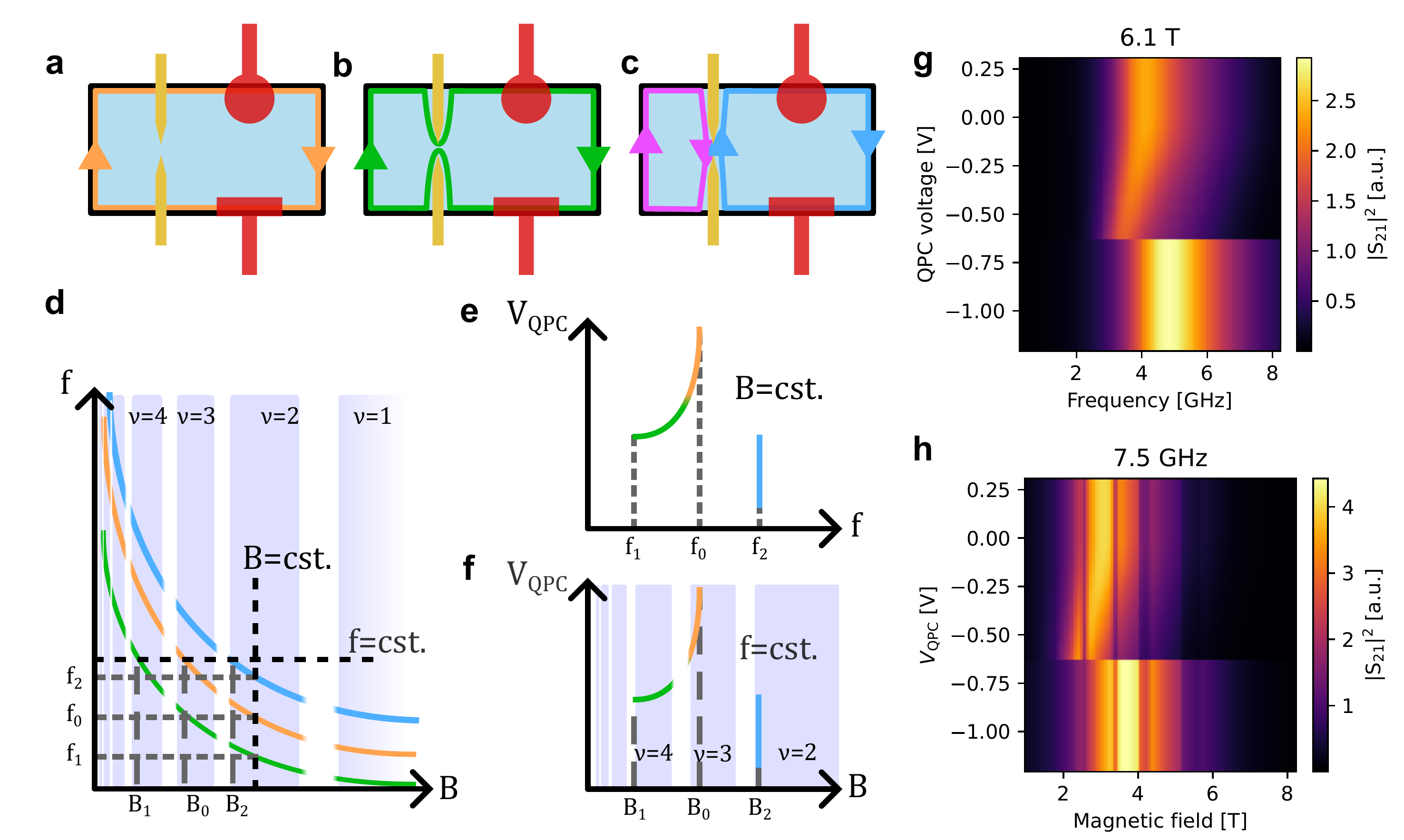}
    \caption{\textbf{Effect of the QPC closing on the EMP resonance:} By applying a voltage on a QPC within an EMP resonator, we go from the configuration of \textbf{(a)} an edge mode traveling around the edge of the full cavity to a \textbf{(b)} pinched mode at the level of the QPC and finally, to \textbf{(c)} two fully separated cavities for very negative voltages. (\textbf{d}) Frequency- and field-dependence of tracked resonance frequencies:  in the $(f,B)$ plane, the open QPC configuration leads to the resonance shown in orange. As the QPC voltage becomes more negative, the perimeter increase leads to the decrease of the resonance frequency shown here in green. Finally, after the pinch-off point, the perimeter is smaller than the one of the original cavity and the resulting resonance curve is found at lower frequency shown here in blue. This could be observed either at fixed magnetic field \textbf{(e)} where the resonance will first decrease in frequency from $f_0$ to $f_1$ before appearing at lower and constant frequency $f_2$. This experiment can also be performed at fixed frequency (\textbf{f}) where the observed resonance drops continuously from $B_0$ to $B_1$ and then shifts toward stronger magnetic field $B_2$ once the QPC is fully closed. Additionally, as the resonance is shifted in the $(V_\mathrm{QPC},B)$ plane, it will cross the various quantum Hall plateaus and as such we expect to observe dark spots where the edge states are not well defined. \textbf{(g)} Simulated map of the transmission as a function of the QPC voltage and frequency for fixed magnetic field. \textbf{(h)} Simulated map of the transmission as a function of the QPC voltage and magnetic field for fixed frequency. Vertical features appear signaling the presence of the quantum Hall plateaus.}
    \label{fig2_principle}
\end{figure*}

Having access to the full field dependence of the resonances allows for the extraction of the depletion length quantitatively using the following analysis. The EMP velocity can be approximated to $v\simeq \frac{e nd}{\varepsilon a B}$ \cite{johnson2003} where $a$ is the depletion length of the 2DEG in a smooth-edge model of a gated system. Assuming the $1/B$ behavior of the EMP velocity shown in previous works  holds here \cite{talyanskii1990edge, talyanskii1992spectroscopy, talyanskii1994experimental, frigerio2024, tarascio2026compact} we extract the quantity $f_0B$, which is a constant for each one of these maps. This is done with the fit shown by the blue lines on figures \ref{fig_gate_capacitance}.a-d. Plotting this quantity as a function of the previously determined electronic density, we obtain figure \ref{fig_gate_capacitance}.f from which we perform a linear fit. As we have already determined all the other parameters of the system through geometry (see supplementary section A), growth parameters or measurement, we can now directly extract the value of the depletion $a=578 \pm 108$~\SI{}{\nano\meter}. 

This value is significantly smaller than the one used in our previous work \cite{frigerio2024} where the plateaus of the quantum Hall effect could not be reliably determined. This has led to an incorrect value of $a$ used in our simulations in this previous work. The value found here is however in very good agreement with previous works on the GaAs based 2DEGs \cite{choi1987experimental, chklovskii1992electrostatics, gelfand1994edge, dahl1995edge}.

\subsection{Splitting of the cavity by a QPC}

We now to turn our attention toward the closing of the QPC in the middle-left cavity (QPC$_2$). Our goal is to reach a regime of tunneling between counter-propagating edge states brought together by the closing of QPC 2. The process of interest is schematically represented on figures \ref{fig2_principle}.a-c. As we start from the middle-left cavity, the edge mode (shown in orange on figure \ref{fig2_principle}.a) is pushed toward the center of the sample by the action of the QPC as seen on figure \ref{fig2_principle}.b. The increase of the EMP path is expected to push the resonance toward lower frequency. However, following the full pinch-off of the QPC, the cavity is then split in two distinct regions, leading to two distinct modes (shown in pink and blue on figure \ref{fig2_principle}.c). Out of these two modes, only the one in the central region which forms the "small" cavity configuration is addressed by the input and output gates and, as shown on figure \ref{fig2_principle}.d, the resulting resonance should appear at higher frequency due to the reduced perimeter of the cavity.

\begin{figure*}[t]
    \begin{center}
    \includegraphics[width = \linewidth]{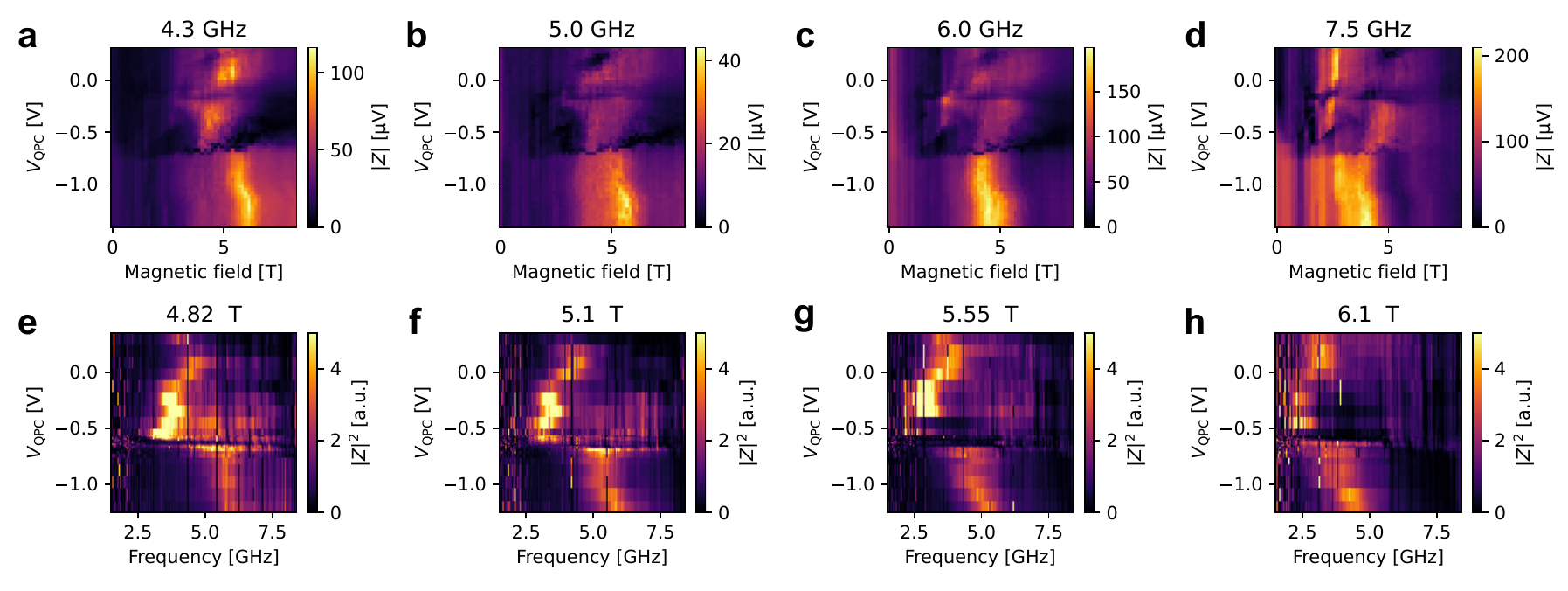}
    \caption{\textbf{Acquisition methods, fixed frequency vs fixed field:} \textbf{(a)-(d)} Transmission signal acquired as a function of gate voltage and magnetic field for frequencies $f=$~\SI{4.3}{\giga\hertz}, \SI{5.0}{\giga\hertz}, \SI{6.0}{\giga\hertz} and \SI{7.5}{\giga\hertz}. The only treatment applied to the data is the subtraction of the background signal. \textbf{(e)-(h)} Transmission signal acquired as a function of QPC voltage and frequency for a magnetic fields $B=$~\SI{4.82}{\tesla}, \SI{5.1}{\tesla}, \SI{5.55}{\tesla} and \SI{6.1}{\tesla}. In order to display the data, the normalization procedure described in \cite{frigerio2024} has been used on every field vs. frequency map before extracting the necessary data to plot this figure. Only the offset has been subtracted from the data.}
    \label{fig4}
    \end{center}
\end{figure*}

There are two different ways to visualize the effect. The first one is to study the dependence of the resonance as a function of QPC voltage and frequency. This has the advantage of focusing on a single magnetic field and, therefore, to work on the same quantum Hall plateau over the whole map. 
However, in order to extract any frequency dependence of the system's response, it is necessary to normalize our transmission maps and to correct for the gain variations as a function of frequency (see \cite{frigerio2024}). Such a procedure necessitates knowing the full magnetic field dependence, leading to an increased acquisition time and therefore a lowered resolution of the experimental data. A simulation of the expected behavior of the cavity is shown in figure 4.e and g, illustrating the gradual frequency shift followed by the abrupt jump of the resonance described above.

The second method is to work at constant frequency, with the advantage of fixing the gain of the detection on a given map. The parasitic contribution to the rf signal that originates from direct coupling between input and output gates and other parasitic capacitive couplings, can then be subtracted simply by simultaneously acquiring the parasitic signal (see supplementary section C). The difference in the final result is that in this configuration, the quantum Hall plateaus will appear, superimposed to the resonance mode as shown on the simulation on figure \ref{fig2_principle}.h. We thus expect the appearance of dark low-signal areas at some specific values of the magnetic fields corresponding to transitions between quantum Hall plateaus. However, as we will show later on the experimental data, this description only applies to a system at fixed electronic density $n_B$. Any change of $n_B$ due to the influence of the QPC will actually lead to a shift of the plateaus and thus of the pattern as a function of $V_\mathrm{QPC}$. In the rest of this paper, the top gates (TG$_1$ and TG$_2$) are polarized with the voltage $V_\mathrm{all~gates} = -75$~\SI{}{\milli\volt}.

The results at fixed frequency are shown on figures \ref{fig4}.a-d where we display the non-normalized transmission amplitude $|Z|$ as a function of the magnetic field and $V_\mathrm{QPC}$. These maps were acquired at 4 different frequencies and in a "constant density" configuration which means that between each two distinct QPC values, we open the outer QPCs in order to allow for the relaxation of charging energy within the cavity. The underlying idea here is that, as QPC 2 closes, the available area for electrons diminishes significantly with they have no way to exit the cavity. This would lead to a constant number of electrons in the cavity over the whole measurement but to an increased density and charging energy as the gate voltage is pushed toward more negative values. Our way of measuring in a "constant density" mode is thus closer to the procedure described on figure \ref{fig2_principle} and, in first approximation, we do not need to invoke a change of the density to explain our findings (see supplementary section C).

Starting at \SI{4.3}{\giga\hertz} (figure \ref{fig4}.a), we observe on the upper part of the transmission map that the resonance shifts as the QPC voltage is decreased. Around \SI{-0.6}{\volt}, we observe a sharp discontinuity that signals the pinch-off point of the QPC. This pinch-off is followed by a shift of the resonance toward higher fields, as expected theoretically. This behavior is observed for all 4 frequencies shown here. However, as the frequency is increased, we start to see another resonance appearing before the pinch-off point and shifting toward lower fields as seen on figures \ref{fig4}.c and d. In particular, a lower intensity area appears between 3 and \SI{4}{\tesla} that we interpret as the transition between the $\nu=1$ and the $\nu = 2$ plateaus.

Another discontinuity is observed around $V_\mathrm{QPC} \simeq -0.2$~\SI{}{\volt} whose origin is not fully understood. Our current hypothesis is that this is linked to the voltage applied on top gates $V_G = -75$~\SI{}{\milli\volt}. Above this specific voltage, the 2DEG below the QPC actually has a higher electronic density than the rest of the resonator, leading to localized states and charging effects caused by localized charges.

For very low values of the QPC voltage, in the small cavity configuration, while we expect a vertical resonance with no additional feature, we instead observe a small shift toward larger fields over a finite range of the QPC voltage. This shift is attributed to electrostatic gating of the cavity by the QPC despite our efforts to work at constant density over the whole voltage range. Reducing the available space for electrons within the cavity increases the electronic density and therefore the velocity of EMPs that lead to a shift of the resonance towards higher fields for a fixed frequency measurement.

On figures \ref{fig4}.e-h, we present another data set obtained in the second configuration where the field is fixed and the frequency is swept. The quantity plotted here is the squared normalized transmission amplitude (see \cite{frigerio2024}).

On these maps, the same behavior is observed of the resonance moving towards lower frequencies as the QPC is pinched-off. The same pinch-point is observed at $V_\mathrm{QPC}\simeq -0.6$~\SI{}{\volt}. Our resolution does not allow us to clearly observe the discontinuity around \SI{0.1}{\volt}. However, the shift of the resonance after the pinch-off point is reproduced in this data set as well. As we explained earlier, because we are working here at constant field for a given map, the filling factor can be taken as constant and we do not observe any signature of the various quantum Hall plateaus.

Close to the pinch-off point of QPC$_2$, we observe multiple resonances shown on figure \ref{fig4}.e. The main resonance, around \SI{3}{\giga\hertz}, can be attributed to the closing "middle left" cavity. The signal around \SI{6}{\giga\hertz} is linked to the first harmonic of the signal that is observable due to the increase of the cavity perimeter near the pinch-off point. A third resonance signal can be observed close to \SI{5}{\giga\hertz} that could originate from the "small" cavity. Indeed, around the pinch-off point, we expect to observe a superposition of a signal from the cavity shown in green on figure \ref{fig2_principle}.b and the one shown in blue on figure \ref{fig2_principle}.c. Such signature was also observed in the "constant number of electrons" configuration and is shown in the supplementary C section of this paper.

\begin{figure}[h]
    \centering
    \includegraphics[width=0.8\linewidth]{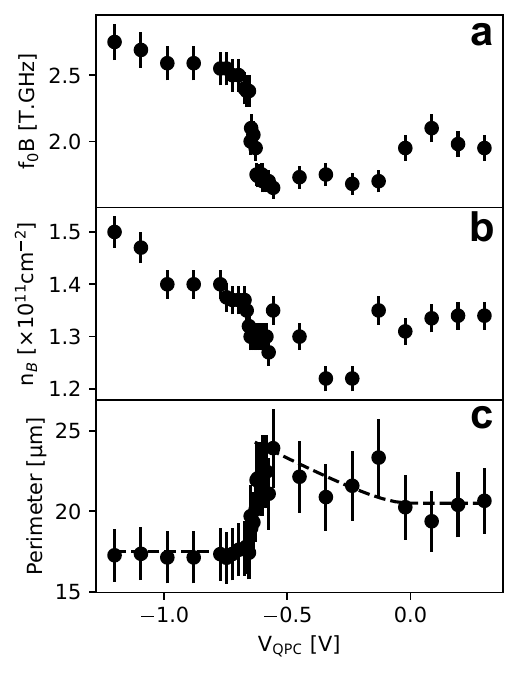}
    \caption{\textbf{Extraction of perimeter:} From the raw data used to plot figures \ref{fig4}.e-h we can extract \textbf{(a)} the factor $f_0B$ by fitting the $1/B$ behavior of the EMP resonance. By identifying the maxima of the signal we then can extract \textbf{(b)} the density $n_B$.  From these two quantities we finally can extract \textbf{(c)} the perimeter of the cavity as a function of $V_\mathrm{QPC}$. The dashed line indicates the geometric perimeter of the cavity estimated from geometric arguments (see appendix) as a function of the gate voltage. A partial data set used to extract these points is shown in the appendix.}
    \label{fig4_value_extraction}
\end{figure}

\subsection{Estimation of cavity perimeter}

We now use the values extracted of the depletion length $a$ and dielectric permittivity $\varepsilon$ to extract the perimeter of the cavity as a function of $V_\mathrm{QPC}$. The only remaining parameter to extract under the action of the QPC are the electronic density and the perimeter of the cavity. By fitting on each $f$ vs. $B$ map (see supplementary section F) the position of the plateaus and the field dependence, similar to those shown on figures \ref{fig_gate_capacitance}.a-d., we are thus able to extract both the electronic density $n_B$ and the constant $f_0B$ as shown on figure \ref{fig4_value_extraction}.a and b from which we then extract the perimeter $P(V_\mathrm{QPC})$ of the cavity.

We observe from these data that, contrary to our expectations, the density actually varies quite drastically with the QPC voltage. As we can see on figure \ref{fig4_value_extraction}, the density increases from $1.3\times 10^{11}$~\SI{}{\centi\meter\squared} at positive voltage to $1.5\times 10^{11}$~\SI{}{\centi\meter\squared} at negative voltage with a drop down to $1.2\times 10^{11}$~\SI{}{\centi\meter\squared} around \SI{-0.25}{\volt}. The sharp increase of the density below \SI{-1}{\volt} in particular explains the shift of the resonance towards larger fields or frequency on figures \ref{fig4}. 

From the knowledge of these two quantities we can directly access the perimeter $P$ (figure \ref{fig4_value_extraction}.c). The  behavior of the perimeter with $V_\mathrm{QPC}$ confirms our expectations detailed on figure \ref{fig2_principle} about the simple picture of an increase of the perimeter followed by a sharp drop at the pinch-off point.

We can see that the perimeters is very stable at \SI{17.5}{\micro\meter} after the pinch-off point which indicates that the increase of density translates into an increase of the velocity (seen on the increase of $f_0B$) and both effects compensate as the EMP path remains unchanged.

In order to compare these results with a simple physical picture, we plot the dashed line on top of the data on figure \ref{fig4_value_extraction}.c. This line is obtained by geometric consideration, assuming that the action of the QPC has the effect of moving the trajectory of EMPs with a linear dependence on the voltage (see supplementary section A). The constraint on this calculation is that for a fully open QPC ($V_\mathrm{QPC}>0$) we should recover the geometric perimeter of the "middle left" cavity.

\subsection{Signature of fractional quantum Hall plateaus in rf}

\begin{figure}[h!]
    \centering
    \includegraphics[width=\linewidth]{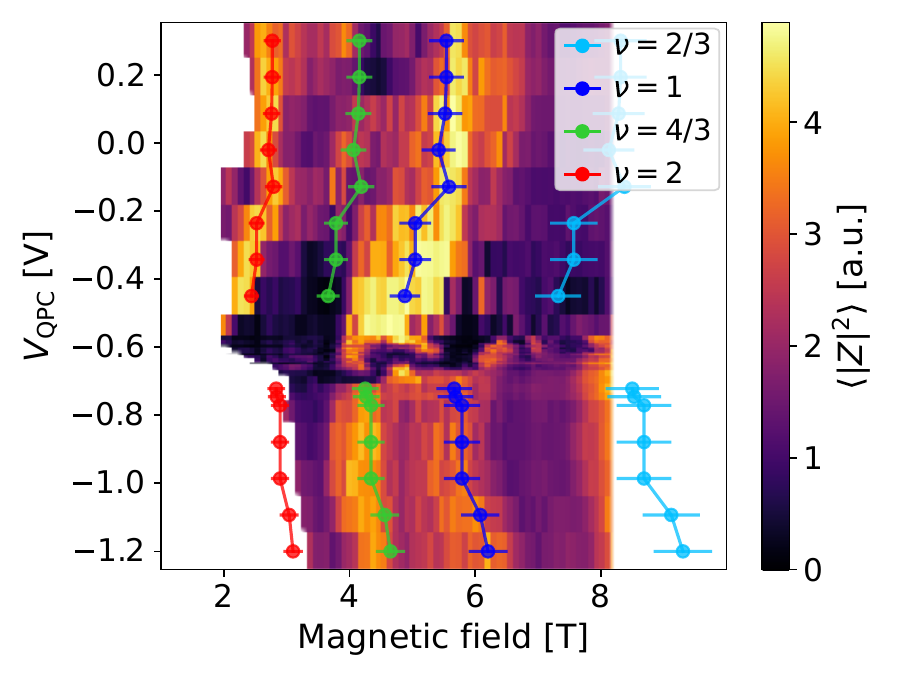}
    \caption{\textbf{Integrated amplitude along resonance:} From full $(f,B)$ transmission maps we can fit the $1/B$ decrease of the resonance frequency for each value of the QPC voltage. By integrating over a \SI{200}{\mega\hertz} frequency window around the central resonance frequency we obtain the following figure that show the resonances associated to the various quantum Hall plateaus as a function of the QPC voltage and magnetic field. For each value of $V_\mathrm{QPC}$ we extract the position of the integer plateaus ($\nu = 2$ for $V_\mathrm{QPC}>-0.5$ and $\nu = 1$ for $V_\mathrm{QPC}<-0.6$) and extrapolate to the expected position of other integer and fractional plateaus shown with dot markers.}
    \label{integrated amplitude}
\end{figure}

In order to obtain a clearer signature of fractional plateaus, we use the previous data shown on figure \ref{fig4_value_extraction}.a of the dependence of $f_0B$ with $V_\mathrm{QPC}$. From these data points we can trace the field dependence of the resonance frequency (like those plotted in light blue on figures \ref{fig_gate_capacitance}.a-d) and study the amplitude of the transmission signal $|Z|^2$ along this line (see appendix F for more details). In practice, for each value of $V_\mathrm{QPC}$ and $B$ we average the $|Z|^2$ signal over a frequency window $\Delta f$ around each point of the hyperbolic line in the $(f_0, B)$ plane. Due to the error of $7.5\%$ on the evaluation of $f_0B$ we have for a fixed magnetic field $\Delta f = 0.075\times f$. The resulting quantity $\left\langle|Z|^2\right\rangle$ is shown on figure \ref{integrated amplitude}. On this two-dimensional map, we observe clear peaks that signal the presence of quantum Hall plateaus. As we described previously these peaks move with $V_\mathrm{QPC}$ until the pinch-off point around \SI{-0.6}{\volt}. After this point the resonance only shifts slightly just as we described in relation to figure \ref{fig4}. Together with the two-dimensional map, we have added the expected position of plateaus from the fit used to obtain figure \ref{fig4_value_extraction}.b. These points align very closely with the position of the observed peaks.

Before the pinch-off point, the two main resonances that are observed are related to the $\nu = 1$ and $\nu = 2$ integer quantum Hall plateaus. Between these we also observe signal associated to the fractional $\nu =4/3$ filling. However, neither the $\nu = 2/3$ nor the $\nu = 5/3$ states are not observed here. The situation is slightly different after the pinch-off point where the $\nu = 2$ plateau is out of our range of measurement (the corresponding resonance shows up for frequency larger than \SI{8}{\giga\hertz}). Contrary to the open QPC configuration, we do observe a strong signal associated to the $\nu = 4/3$ filling, unequivocally signaling the presence of a fractional quantum Hall states. At larger magnetic fields we also observe the tail of the $\nu = 2/3$ associated resonance that is unfortunately centered around a larger magnetic field which was not accessible in our experimental setup. Nevertheless, because our resonances are intrinsically large we do manage to observe the edge of the resonance around \SI{8}{\tesla}.

The reason why some plateaus appear clearer than others is not currently understood but seems to be linked to the state and geometry of the QPC as these signatures are better seen in a "small cavity" configuration in which no QPC is included within the EMP cavity. This observation suggests that in future experiments, the geometry of QPCs should be carefully designed and optimized in order to maximize the transmission signal associated to fractional states if we are to attempt studying anyonic signatures in an EMP rf interferometer.

At the level of the pinch-off point where we have a larger voltage resolution (for $-0.7~\mathrm{V}<V_\mathrm{QPC}<-0.6~\mathrm{V}$) we observe a splitting of the resonance peaks. This effects is most clearly seen on the $\nu = 1$ state. This could be interpreted as the coexistence of two distinct paths for EMPs, one that corresponds to the small cavity (in blue in figure \ref{fig2_principle}.b) as EMPs get back-scattered at the level of the QPC and one that corresponds to the transmitted EMPs that explore the full "middle left" cavity and go around the pinching QPCs (in green on figure \ref{fig2_principle}.b). This interpretation would then indicate the possibility to have the coexistence of multiple modes at the same filling factor with different paths, thus paving the way towards the realization of an rf EMP interferometer. However, more work will be needed to confirm this interpretation of our results.

\section{Conclusion}

In conclusion, we have studied an EMP resonator defined by electrostatic gating. Using the electrostatic gates we were able to determine the value of the dielectric permittivity of the AlGaAs layer separating the top gate from the 2DEG as well as the depletion length of the 2DEG at the edge of the sample. By comparing dc and rf data we were able to clearly identify rf resonances associated to fractional quantum Hall states $\nu =4/3$ and $2/3$. We followed this analysis by studying the closing of a QPC within the radiofrequency resonator defined by confining EMPs. We have studied the behavior of the cavity as a function of magnetic field and frequency and provided a conceptual framework to understand our experimental results. Ultimately, we managed to extract the perimeter of the cavity by extracting the electronic density and EMP velocity independently, and to model it quantitatively by using a simple geometrical model of the QPCs as a function of the applied gate voltage. Following the $1/B$ trend of the resonance we also present the evolution of the amplitude of the transmission signal with magnetic field and $V_\mathrm{QPC}$ that further highlights the signature of fractional Hall states $\nu = 4/3$ and $2/3$.

While we do not observe any interferometric signature due to the very small expected period of Aharonov-Bohm oscillations, we manage to benchmark the behavior of the system for future investigation and observe the signature of multiple cavities near the pinch-off point. Building on this knowledge will allow further optimization and investigation in these EMP resonators opening the possibility to observe clear signatures of interferometric behavior.

\section{Acknowledgments}

This project received funding from the ANR grant ‘MAHR’, reference no. ANR-24-CE47-3746 (G.C.M.). This project was also funded within the Quantera II Program that has received funding for the ElQURes project from the EU H2020 research and innovation program under the GA No 101017733, and with funding organization ANR, reference ANR-24-QUA2-0002 (G.C.M.). This study benefited from the support the C2N technological facilities, member of the French RENATECH network. We have applied a CC-BY public copyright license to any Author Accepted Manuscript (AAM) version arising from this submission. The authors would like to thank Niels Ubbelohde and Vyacheslavs Kashcheyevs for useful discussions and feedback during the writing of this paper.

\section{Author contribution}

G.C.M. devised the experiment. S.K., E.F., S.P. and G.C.M. performed the experiment and analyzed the data. U.G. and A.C. grew the heterostructure. E.F. and G.C.M designed the samples. Y.J. Fabricated the samples. G.R. and P.D. devised the theoretical description of the system. G.C.M. wrote the paper with the help of S.K., S.P., E.F., G.F., F.D.P., G.R, P.D., J.-M.B. and E.B. All authors have read and commented on the content of the paper.

\bibliography{biblio}

\begin{thebibliography}{31}%
\makeatletter
\providecommand \@ifxundefined [1]{%
 \@ifx{#1\undefined}
}%
\providecommand \@ifnum [1]{%
 \ifnum #1\expandafter \@firstoftwo
 \else \expandafter \@secondoftwo
 \fi
}%
\providecommand \@ifx [1]{%
 \ifx #1\expandafter \@firstoftwo
 \else \expandafter \@secondoftwo
 \fi
}%
\providecommand \natexlab [1]{#1}%
\providecommand \enquote  [1]{``#1''}%
\providecommand \bibnamefont  [1]{#1}%
\providecommand \bibfnamefont [1]{#1}%
\providecommand \citenamefont [1]{#1}%
\providecommand \href@noop [0]{\@secondoftwo}%
\providecommand \href [0]{\begingroup \@sanitize@url \@href}%
\providecommand \@href[1]{\@@startlink{#1}\@@href}%
\providecommand \@@href[1]{\endgroup#1\@@endlink}%
\providecommand \@sanitize@url [0]{\catcode `\\12\catcode `\$12\catcode `\&12\catcode `\#12\catcode `\^12\catcode `\_12\catcode `\%12\relax}%
\providecommand \@@startlink[1]{}%
\providecommand \@@endlink[0]{}%
\providecommand \url  [0]{\begingroup\@sanitize@url \@url }%
\providecommand \@url [1]{\endgroup\@href {#1}{\urlprefix }}%
\providecommand \urlprefix  [0]{URL }%
\providecommand \Eprint [0]{\href }%
\providecommand \doibase [0]{https://doi.org/}%
\providecommand \selectlanguage [0]{\@gobble}%
\providecommand \bibinfo  [0]{\@secondoftwo}%
\providecommand \bibfield  [0]{\@secondoftwo}%
\providecommand \translation [1]{[#1]}%
\providecommand \BibitemOpen [0]{}%
\providecommand \bibitemStop [0]{}%
\providecommand \bibitemNoStop [0]{.\EOS\space}%
\providecommand \EOS [0]{\spacefactor3000\relax}%
\providecommand \BibitemShut  [1]{\csname bibitem#1\endcsname}%
\let\auto@bib@innerbib\@empty
\bibitem [{\citenamefont {Andrei}\ \emph {et~al.}(1988)\citenamefont {Andrei}, \citenamefont {Glattli}, \citenamefont {Williams},\ and\ \citenamefont {Heiblum}}]{andrei1988low}%
  \BibitemOpen
  \bibfield  {author} {\bibinfo {author} {\bibfnamefont {E.}~\bibnamefont {Andrei}}, \bibinfo {author} {\bibfnamefont {D.}~\bibnamefont {Glattli}}, \bibinfo {author} {\bibfnamefont {F.}~\bibnamefont {Williams}},\ and\ \bibinfo {author} {\bibfnamefont {M.}~\bibnamefont {Heiblum}},\ }\bibfield  {title} {\bibinfo {title} {Low frequency collective excitations in the quantum-{H}all system},\ }\href {https://www.sciencedirect.com/science/article/pii/0039602888907327} {\bibfield  {journal} {\bibinfo  {journal} {Surface science}\ }\textbf {\bibinfo {volume} {196}},\ \bibinfo {pages} {501} (\bibinfo {year} {1988})}\BibitemShut {NoStop}%
\bibitem [{\citenamefont {Volkov}\ and\ \citenamefont {Mikhailov}(1988)}]{volkov1988edge}%
  \BibitemOpen
  \bibfield  {author} {\bibinfo {author} {\bibfnamefont {V.~A.}\ \bibnamefont {Volkov}}\ and\ \bibinfo {author} {\bibfnamefont {S.~A.}\ \bibnamefont {Mikhailov}},\ }\bibfield  {title} {\bibinfo {title} {Edge magnetoplasmons: low frequency weakly damped excitations in inhomogeneous two-dimensional electron systems},\ }\href {https://jetp.ras.ru/cgi-bin/e/index/e/67/8/p1639?a=list} {\bibfield  {journal} {\bibinfo  {journal} {Sov. Phys. JETP}\ }\textbf {\bibinfo {volume} {67}},\ \bibinfo {pages} {1639} (\bibinfo {year} {1988})}\BibitemShut {NoStop}%
\bibitem [{\citenamefont {Safi}\ and\ \citenamefont {Schulz}(1995)}]{safi1995}%
  \BibitemOpen
  \bibfield  {author} {\bibinfo {author} {\bibfnamefont {I.}~\bibnamefont {Safi}}\ and\ \bibinfo {author} {\bibfnamefont {H.}~\bibnamefont {Schulz}},\ }\bibfield  {title} {\bibinfo {title} {Transport in an inhomogeneous interacting one-dimensional system},\ }\href {https://journals.aps.org/prb/abstract/10.1103/PhysRevB.52.R17040} {\bibfield  {journal} {\bibinfo  {journal} {Physical Review B}\ }\textbf {\bibinfo {volume} {52}},\ \bibinfo {pages} {R17040} (\bibinfo {year} {1995})}\BibitemShut {NoStop}%
\bibitem [{\citenamefont {Von~Delft}\ and\ \citenamefont {Schoeller}(1998)}]{von1998}%
  \BibitemOpen
  \bibfield  {author} {\bibinfo {author} {\bibfnamefont {J.}~\bibnamefont {Von~Delft}}\ and\ \bibinfo {author} {\bibfnamefont {H.}~\bibnamefont {Schoeller}},\ }\bibfield  {title} {\bibinfo {title} {Bosonization for beginners—refermionization for experts},\ }\href {https://onlinelibrary.wiley.com/doi/10.1002/andp.19985100401} {\bibfield  {journal} {\bibinfo  {journal} {Annalen der Physik}\ }\textbf {\bibinfo {volume} {510}},\ \bibinfo {pages} {225} (\bibinfo {year} {1998})}\BibitemShut {NoStop}%
\bibitem [{\citenamefont {Safi}(1999)}]{safi1999}%
  \BibitemOpen
  \bibfield  {author} {\bibinfo {author} {\bibfnamefont {I.}~\bibnamefont {Safi}},\ }\bibfield  {title} {\bibinfo {title} {A dynamic scattering approach for a gated interacting wire},\ }\href {https://link.springer.com/article/10.1007/s100510051026} {\bibfield  {journal} {\bibinfo  {journal} {Eur. Phys. J}\ }\textbf {\bibinfo {volume} {12}},\ \bibinfo {pages} {451} (\bibinfo {year} {1999})}\BibitemShut {NoStop}%
\bibitem [{\citenamefont {S\'en\'echal}(1999)}]{senechal1999}%
  \BibitemOpen
  \bibfield  {author} {\bibinfo {author} {\bibfnamefont {D.}~\bibnamefont {S\'en\'echal}},\ }\bibfield  {title} {\bibinfo {title} {An introduction to bosonization, eprint},\ }\href {https://arxiv.org/abs/cond-mat/9908262} {\bibfield  {journal} {\bibinfo  {journal} {arXiv preprint cond-mat/9908262}\ } (\bibinfo {year} {1999})}\BibitemShut {NoStop}%
\bibitem [{\citenamefont {Talyanskii}\ \emph {et~al.}(1990)\citenamefont {Talyanskii}, \citenamefont {Wassermeier}, \citenamefont {Wixforth}, \citenamefont {Oshinowo}, \citenamefont {Kotthaus}, \citenamefont {Batov}, \citenamefont {Weimann}, \citenamefont {Nickel},\ and\ \citenamefont {Schlapp}}]{talyanskii1990edge}%
  \BibitemOpen
  \bibfield  {author} {\bibinfo {author} {\bibfnamefont {V.~K.}\ \bibnamefont {Talyanskii}}, \bibinfo {author} {\bibfnamefont {M.}~\bibnamefont {Wassermeier}}, \bibinfo {author} {\bibfnamefont {A.}~\bibnamefont {Wixforth}}, \bibinfo {author} {\bibfnamefont {J.}~\bibnamefont {Oshinowo}}, \bibinfo {author} {\bibfnamefont {J.~P.}\ \bibnamefont {Kotthaus}}, \bibinfo {author} {\bibfnamefont {I.~E.}\ \bibnamefont {Batov}}, \bibinfo {author} {\bibfnamefont {G.}~\bibnamefont {Weimann}}, \bibinfo {author} {\bibfnamefont {H.}~\bibnamefont {Nickel}},\ and\ \bibinfo {author} {\bibfnamefont {W.}~\bibnamefont {Schlapp}},\ }\bibfield  {title} {\bibinfo {title} {Edge magnetoplasmons in the quantum {H}all effect regime},\ }\href {https://www.sciencedirect.com/science/article/pii/003960289090827U} {\bibfield  {journal} {\bibinfo  {journal} {Surface science}\ }\textbf {\bibinfo {volume} {229}},\ \bibinfo {pages} {40} (\bibinfo {year} {1990})}\BibitemShut {NoStop}%
\bibitem [{\citenamefont {Talyanskii}\ \emph {et~al.}(1992)\citenamefont {Talyanskii}, \citenamefont {Polisski}, \citenamefont {Arnone}, \citenamefont {Pepper}, \citenamefont {Smith}, \citenamefont {Ritchie}, \citenamefont {Frost},\ and\ \citenamefont {Jones}}]{talyanskii1992spectroscopy}%
  \BibitemOpen
  \bibfield  {author} {\bibinfo {author} {\bibfnamefont {V.}~\bibnamefont {Talyanskii}}, \bibinfo {author} {\bibfnamefont {A.}~\bibnamefont {Polisski}}, \bibinfo {author} {\bibfnamefont {D.}~\bibnamefont {Arnone}}, \bibinfo {author} {\bibfnamefont {M.}~\bibnamefont {Pepper}}, \bibinfo {author} {\bibfnamefont {C.}~\bibnamefont {Smith}}, \bibinfo {author} {\bibfnamefont {D.}~\bibnamefont {Ritchie}}, \bibinfo {author} {\bibfnamefont {J.}~\bibnamefont {Frost}},\ and\ \bibinfo {author} {\bibfnamefont {G.}~\bibnamefont {Jones}},\ }\bibfield  {title} {\bibinfo {title} {Spectroscopy of a two-dimensional electron gas in the quantum-{H}all-effect regime by use of low-frequency edge magnetoplasmons},\ }\href {https://journals.aps.org/prb/abstract/10.1103/PhysRevB.46.12427} {\bibfield  {journal} {\bibinfo  {journal} {Physical Review B}\ }\textbf {\bibinfo {volume} {46}},\ \bibinfo {pages} {12427} (\bibinfo {year} {1992})}\BibitemShut {NoStop}%
\bibitem [{\citenamefont {Balaban}\ \emph {et~al.}(1997)\citenamefont {Balaban}, \citenamefont {Meirav}, \citenamefont {Shtrikman},\ and\ \citenamefont {Umansky}}]{balaban1997observation}%
  \BibitemOpen
  \bibfield  {author} {\bibinfo {author} {\bibfnamefont {N.}~\bibnamefont {Balaban}}, \bibinfo {author} {\bibfnamefont {U.}~\bibnamefont {Meirav}}, \bibinfo {author} {\bibfnamefont {H.}~\bibnamefont {Shtrikman}},\ and\ \bibinfo {author} {\bibfnamefont {V.}~\bibnamefont {Umansky}},\ }\bibfield  {title} {\bibinfo {title} {Observation of the logarithmic dispersion of high-frequency edge excitations},\ }\href {https://harvest.aps.org/v2/journals/articles/10.1103/PhysRevB.55.R13397/fulltext} {\bibfield  {journal} {\bibinfo  {journal} {Physical Review B}\ }\textbf {\bibinfo {volume} {55}},\ \bibinfo {pages} {R13397} (\bibinfo {year} {1997})}\BibitemShut {NoStop}%
\bibitem [{\citenamefont {Hashisaka}\ \emph {et~al.}(2013)\citenamefont {Hashisaka}, \citenamefont {Kamata}, \citenamefont {Kumada}, \citenamefont {Washio}, \citenamefont {Murata}, \citenamefont {Muraki},\ and\ \citenamefont {Fujisawa}}]{hashisaka2013}%
  \BibitemOpen
  \bibfield  {author} {\bibinfo {author} {\bibfnamefont {M.}~\bibnamefont {Hashisaka}}, \bibinfo {author} {\bibfnamefont {H.}~\bibnamefont {Kamata}}, \bibinfo {author} {\bibfnamefont {N.}~\bibnamefont {Kumada}}, \bibinfo {author} {\bibfnamefont {K.}~\bibnamefont {Washio}}, \bibinfo {author} {\bibfnamefont {R.}~\bibnamefont {Murata}}, \bibinfo {author} {\bibfnamefont {K.}~\bibnamefont {Muraki}},\ and\ \bibinfo {author} {\bibfnamefont {T.}~\bibnamefont {Fujisawa}},\ }\bibfield  {title} {\bibinfo {title} {Distributed-element circuit model of edge magnetoplasmon transport},\ }\href {https://journals.aps.org/prb/abstract/10.1103/PhysRevB.88.235409} {\bibfield  {journal} {\bibinfo  {journal} {Physical Review B}\ }\textbf {\bibinfo {volume} {88}},\ \bibinfo {pages} {235409} (\bibinfo {year} {2013})}\BibitemShut {NoStop}%
\bibitem [{\citenamefont {Kumada}\ \emph {et~al.}(2014)\citenamefont {Kumada}, \citenamefont {Roulleau}, \citenamefont {Roche}, \citenamefont {Hashisaka}, \citenamefont {Hibino}, \citenamefont {Petkovi{\'c}},\ and\ \citenamefont {Glattli}}]{kumada2014resonant}%
  \BibitemOpen
  \bibfield  {author} {\bibinfo {author} {\bibfnamefont {N.}~\bibnamefont {Kumada}}, \bibinfo {author} {\bibfnamefont {P.}~\bibnamefont {Roulleau}}, \bibinfo {author} {\bibfnamefont {P.}~\bibnamefont {Roche}}, \bibinfo {author} {\bibfnamefont {M.}~\bibnamefont {Hashisaka}}, \bibinfo {author} {\bibfnamefont {H.}~\bibnamefont {Hibino}}, \bibinfo {author} {\bibfnamefont {I.}~\bibnamefont {Petkovi{\'c}}},\ and\ \bibinfo {author} {\bibfnamefont {D.}~\bibnamefont {Glattli}},\ }\bibfield  {title} {\bibinfo {title} {Resonant edge magnetoplasmons and their decay in graphene},\ }\href {https://journals.aps.org/prl/abstract/10.1103/PhysRevLett.113.266601} {\bibfield  {journal} {\bibinfo  {journal} {Physical review letters}\ }\textbf {\bibinfo {volume} {113}},\ \bibinfo {pages} {266601} (\bibinfo {year} {2014})}\BibitemShut {NoStop}%
\bibitem [{\citenamefont {Lin}\ \emph {et~al.}(2024)\citenamefont {Lin}, \citenamefont {Futamata}, \citenamefont {Akiho}, \citenamefont {Muraki},\ and\ \citenamefont {Fujisawa}}]{lin2024resonant}%
  \BibitemOpen
  \bibfield  {author} {\bibinfo {author} {\bibfnamefont {C.}~\bibnamefont {Lin}}, \bibinfo {author} {\bibfnamefont {K.}~\bibnamefont {Futamata}}, \bibinfo {author} {\bibfnamefont {T.}~\bibnamefont {Akiho}}, \bibinfo {author} {\bibfnamefont {K.}~\bibnamefont {Muraki}},\ and\ \bibinfo {author} {\bibfnamefont {T.}~\bibnamefont {Fujisawa}},\ }\bibfield  {title} {\bibinfo {title} {Resonant plasmon-assisted tunneling in a double quantum dot coupled to a quantum {H}all plasmon resonator},\ }\href {https://journals.aps.org/prl/abstract/10.1103/PhysRevLett.133.036301} {\bibfield  {journal} {\bibinfo  {journal} {Physical Review Letters}\ }\textbf {\bibinfo {volume} {133}},\ \bibinfo {pages} {036301} (\bibinfo {year} {2024})}\BibitemShut {NoStop}%
\bibitem [{\citenamefont {Lin}\ \emph {et~al.}(2026)\citenamefont {Lin}, \citenamefont {Teshima}, \citenamefont {Akiho}, \citenamefont {Muraki},\ and\ \citenamefont {Fujisawa}}]{lin2026dispersive}%
  \BibitemOpen
  \bibfield  {author} {\bibinfo {author} {\bibfnamefont {C.}~\bibnamefont {Lin}}, \bibinfo {author} {\bibfnamefont {K.}~\bibnamefont {Teshima}}, \bibinfo {author} {\bibfnamefont {T.}~\bibnamefont {Akiho}}, \bibinfo {author} {\bibfnamefont {K.}~\bibnamefont {Muraki}},\ and\ \bibinfo {author} {\bibfnamefont {T.}~\bibnamefont {Fujisawa}},\ }\bibfield  {title} {\bibinfo {title} {Dispersive detection of a charge qubit with a broadband high-impedance quantum-hall plasmon resonator},\ }\href {https://www.nature.com/articles/s41467-026-69342-y} {\bibfield  {journal} {\bibinfo  {journal} {Nature Communications}\ }\textbf {\bibinfo {volume} {17}},\ \bibinfo {pages} {2600} (\bibinfo {year} {2026})}\BibitemShut {NoStop}%
\bibitem [{\citenamefont {Bosco}\ and\ \citenamefont {Di~Vicenzo}(2019)}]{bosco2019}%
  \BibitemOpen
  \bibfield  {author} {\bibinfo {author} {\bibfnamefont {S.}~\bibnamefont {Bosco}}\ and\ \bibinfo {author} {\bibfnamefont {D.~P.}\ \bibnamefont {Di~Vicenzo}},\ }\bibfield  {title} {\bibinfo {title} {Transmission lines and resonators based on quantum {H}all plasmonics: {E}lectromagnetic field, attenuation, and coupling to qubits},\ }\href {https://journals.aps.org/prb/abstract/10.1103/PhysRevB.100.035416} {\bibfield  {journal} {\bibinfo  {journal} {Physical Review B}\ }\textbf {\bibinfo {volume} {100}},\ \bibinfo {pages} {035416} (\bibinfo {year} {2019})}\BibitemShut {NoStop}%
\bibitem [{\citenamefont {Mahoney}\ \emph {et~al.}(2017)\citenamefont {Mahoney}, \citenamefont {Colless}, \citenamefont {Pauka}, \citenamefont {Hornibrook}, \citenamefont {Watson}, \citenamefont {Gardner}, \citenamefont {Manfra}, \citenamefont {Doherty},\ and\ \citenamefont {Reilly}}]{mahoney2017chip}%
  \BibitemOpen
  \bibfield  {author} {\bibinfo {author} {\bibfnamefont {A.}~\bibnamefont {Mahoney}}, \bibinfo {author} {\bibfnamefont {J.}~\bibnamefont {Colless}}, \bibinfo {author} {\bibfnamefont {S.}~\bibnamefont {Pauka}}, \bibinfo {author} {\bibfnamefont {J.}~\bibnamefont {Hornibrook}}, \bibinfo {author} {\bibfnamefont {J.}~\bibnamefont {Watson}}, \bibinfo {author} {\bibfnamefont {G.}~\bibnamefont {Gardner}}, \bibinfo {author} {\bibfnamefont {M.}~\bibnamefont {Manfra}}, \bibinfo {author} {\bibfnamefont {A.}~\bibnamefont {Doherty}},\ and\ \bibinfo {author} {\bibfnamefont {D.}~\bibnamefont {Reilly}},\ }\bibfield  {title} {\bibinfo {title} {On-chip microwave quantum {H}all circulator},\ }\href {https://journals.aps.org/prx/abstract/10.1103/PhysRevX.7.011007} {\bibfield  {journal} {\bibinfo  {journal} {Physical Review X}\ }\textbf {\bibinfo {volume} {7}},\ \bibinfo {pages} {011007} (\bibinfo {year} {2017})}\BibitemShut {NoStop}%
\bibitem [{\citenamefont {Martinez}\ \emph {et~al.}(2024)\citenamefont {Martinez}, \citenamefont {Qiu}, \citenamefont {Deng}, \citenamefont {Zhang}, \citenamefont {Ray}, \citenamefont {Tai}, \citenamefont {Wei}, \citenamefont {He}, \citenamefont {Wang}, \citenamefont {DuBois},\ and\ \citenamefont {Qu}}]{martinez2024edge}%
  \BibitemOpen
  \bibfield  {author} {\bibinfo {author} {\bibfnamefont {L.~A.}\ \bibnamefont {Martinez}}, \bibinfo {author} {\bibfnamefont {G.}~\bibnamefont {Qiu}}, \bibinfo {author} {\bibfnamefont {P.}~\bibnamefont {Deng}}, \bibinfo {author} {\bibfnamefont {P.}~\bibnamefont {Zhang}}, \bibinfo {author} {\bibfnamefont {K.~G.}\ \bibnamefont {Ray}}, \bibinfo {author} {\bibfnamefont {L.}~\bibnamefont {Tai}}, \bibinfo {author} {\bibfnamefont {M.-T.}\ \bibnamefont {Wei}}, \bibinfo {author} {\bibfnamefont {H.}~\bibnamefont {He}}, \bibinfo {author} {\bibfnamefont {K.~L.}\ \bibnamefont {Wang}}, \bibinfo {author} {\bibfnamefont {J.~L.}\ \bibnamefont {DuBois}},\ and\ \bibinfo {author} {\bibfnamefont {D.-X.}\ \bibnamefont {Qu}},\ }\bibfield  {title} {\bibinfo {title} {Edge magnetoplasmon dispersion and time-resolved plasmon transport in a quantum anomalous {H}all insulator},\ }\href {https://journals.aps.org/prresearch/abstract/10.1103/PhysRevResearch.6.013081} {\bibfield  {journal} {\bibinfo  {journal} {Physical Review Research}\
  }\textbf {\bibinfo {volume} {6}},\ \bibinfo {pages} {013081} (\bibinfo {year} {2024})}\BibitemShut {NoStop}%
\bibitem [{\citenamefont {Tarascio}\ \emph {et~al.}(2026)\citenamefont {Tarascio}, \citenamefont {Zhao}, \citenamefont {Eggli}, \citenamefont {Patlatiuk}, \citenamefont {Reichl}, \citenamefont {Wegscheider}, \citenamefont {Bosco},\ and\ \citenamefont {Zumb{\"u}hl}}]{tarascio2026compact}%
  \BibitemOpen
  \bibfield  {author} {\bibinfo {author} {\bibfnamefont {A.}~\bibnamefont {Tarascio}}, \bibinfo {author} {\bibfnamefont {Y.}~\bibnamefont {Zhao}}, \bibinfo {author} {\bibfnamefont {R.~S.}\ \bibnamefont {Eggli}}, \bibinfo {author} {\bibfnamefont {T.}~\bibnamefont {Patlatiuk}}, \bibinfo {author} {\bibfnamefont {C.}~\bibnamefont {Reichl}}, \bibinfo {author} {\bibfnamefont {W.}~\bibnamefont {Wegscheider}}, \bibinfo {author} {\bibfnamefont {S.}~\bibnamefont {Bosco}},\ and\ \bibinfo {author} {\bibfnamefont {D.~M.}\ \bibnamefont {Zumb{\"u}hl}},\ }\bibfield  {title} {\bibinfo {title} {Compact self-matched gyrators using edge magnetoplasmons},\ }\href {https://arxiv.org/abs/2602.05439} {\bibfield  {journal} {\bibinfo  {journal} {arXiv preprint arXiv:2602.05439}\ } (\bibinfo {year} {2026})}\BibitemShut {NoStop}%
\bibitem [{\citenamefont {Bosco}\ \emph {et~al.}(2017)\citenamefont {Bosco}, \citenamefont {Haupt},\ and\ \citenamefont {DiVincenzo}}]{bosco2017self}%
  \BibitemOpen
  \bibfield  {author} {\bibinfo {author} {\bibfnamefont {S.}~\bibnamefont {Bosco}}, \bibinfo {author} {\bibfnamefont {F.}~\bibnamefont {Haupt}},\ and\ \bibinfo {author} {\bibfnamefont {D.~P.}\ \bibnamefont {DiVincenzo}},\ }\bibfield  {title} {\bibinfo {title} {Self-impedance-matched {H}all-effect gyrators and circulators},\ }\href {https://journals.aps.org/prapplied/abstract/10.1103/PhysRevApplied.7.024030} {\bibfield  {journal} {\bibinfo  {journal} {Physical review applied}\ }\textbf {\bibinfo {volume} {7}},\ \bibinfo {pages} {024030} (\bibinfo {year} {2017})}\BibitemShut {NoStop}%
\bibitem [{\citenamefont {Fijalkowski}\ \emph {et~al.}(2024)\citenamefont {Fijalkowski}, \citenamefont {Liu}, \citenamefont {Klement}, \citenamefont {Schreyeck}, \citenamefont {Brunner}, \citenamefont {Gould},\ and\ \citenamefont {Molenkamp}}]{fijalkowski2024balanced}%
  \BibitemOpen
  \bibfield  {author} {\bibinfo {author} {\bibfnamefont {K.~M.}\ \bibnamefont {Fijalkowski}}, \bibinfo {author} {\bibfnamefont {N.}~\bibnamefont {Liu}}, \bibinfo {author} {\bibfnamefont {M.}~\bibnamefont {Klement}}, \bibinfo {author} {\bibfnamefont {S.}~\bibnamefont {Schreyeck}}, \bibinfo {author} {\bibfnamefont {K.}~\bibnamefont {Brunner}}, \bibinfo {author} {\bibfnamefont {C.}~\bibnamefont {Gould}},\ and\ \bibinfo {author} {\bibfnamefont {L.~W.}\ \bibnamefont {Molenkamp}},\ }\bibfield  {title} {\bibinfo {title} {A balanced quantum {H}all resistor},\ }\href {https://www.nature.com/articles/s41928-024-01156-6} {\bibfield  {journal} {\bibinfo  {journal} {Nature Electronics}\ }\textbf {\bibinfo {volume} {7}},\ \bibinfo {pages} {438} (\bibinfo {year} {2024})}\BibitemShut {NoStop}%
\bibitem [{\citenamefont {Frigerio}\ \emph {et~al.}(2024)\citenamefont {Frigerio}, \citenamefont {Rebora}, \citenamefont {Ruelle}, \citenamefont {Souquet-Basi{\`e}ge}, \citenamefont {Jin}, \citenamefont {Gennser}, \citenamefont {Cavanna}, \citenamefont {Pla{\c{c}}ais}, \citenamefont {Baudin}, \citenamefont {Berroir}, \citenamefont {Safi}, \citenamefont {Degiovanni}, \citenamefont {Fève},\ and\ \citenamefont {Ménard}}]{frigerio2024}%
  \BibitemOpen
  \bibfield  {author} {\bibinfo {author} {\bibfnamefont {E.}~\bibnamefont {Frigerio}}, \bibinfo {author} {\bibfnamefont {G.}~\bibnamefont {Rebora}}, \bibinfo {author} {\bibfnamefont {M.}~\bibnamefont {Ruelle}}, \bibinfo {author} {\bibfnamefont {H.}~\bibnamefont {Souquet-Basi{\`e}ge}}, \bibinfo {author} {\bibfnamefont {Y.}~\bibnamefont {Jin}}, \bibinfo {author} {\bibfnamefont {U.}~\bibnamefont {Gennser}}, \bibinfo {author} {\bibfnamefont {A.}~\bibnamefont {Cavanna}}, \bibinfo {author} {\bibfnamefont {B.}~\bibnamefont {Pla{\c{c}}ais}}, \bibinfo {author} {\bibfnamefont {E.}~\bibnamefont {Baudin}}, \bibinfo {author} {\bibfnamefont {J.-M.}\ \bibnamefont {Berroir}}, \bibinfo {author} {\bibfnamefont {I.}~\bibnamefont {Safi}}, \bibinfo {author} {\bibfnamefont {P.}~\bibnamefont {Degiovanni}}, \bibinfo {author} {\bibfnamefont {G.}~\bibnamefont {Fève}},\ and\ \bibinfo {author} {\bibfnamefont {G.}~\bibnamefont {Ménard}},\ }\bibfield  {title} {\bibinfo {title} {Gate tunable edge magnetoplasmon resonators},\ }\href
  {https://www.nature.com/articles/s42005-024-01803-6} {\bibfield  {journal} {\bibinfo  {journal} {Communications Physics}\ }\textbf {\bibinfo {volume} {7}},\ \bibinfo {pages} {314} (\bibinfo {year} {2024})}\BibitemShut {NoStop}%
\bibitem [{\citenamefont {Cano}\ \emph {et~al.}(2013)\citenamefont {Cano}, \citenamefont {Doherty}, \citenamefont {Nayak},\ and\ \citenamefont {Reilly}}]{cano2013microwave}%
  \BibitemOpen
  \bibfield  {author} {\bibinfo {author} {\bibfnamefont {J.}~\bibnamefont {Cano}}, \bibinfo {author} {\bibfnamefont {A.~C.}\ \bibnamefont {Doherty}}, \bibinfo {author} {\bibfnamefont {C.}~\bibnamefont {Nayak}},\ and\ \bibinfo {author} {\bibfnamefont {D.~J.}\ \bibnamefont {Reilly}},\ }\bibfield  {title} {\bibinfo {title} {Microwave absorption by a mesoscopic quantum {H}all droplet},\ }\href {https://journals.aps.org/prb/abstract/10.1103/PhysRevB.88.165305} {\bibfield  {journal} {\bibinfo  {journal} {Physical Review B}\ }\textbf {\bibinfo {volume} {88}},\ \bibinfo {pages} {165305} (\bibinfo {year} {2013})}\BibitemShut {NoStop}%
\bibitem [{\citenamefont {Murabayashi}\ \emph {et~al.}(2026)\citenamefont {Murabayashi}, \citenamefont {Sano}, \citenamefont {Ronetti}, \citenamefont {Rech}, \citenamefont {Martin}, \citenamefont {Jonckheere},\ and\ \citenamefont {Kato}}]{microwave2025marseille}%
  \BibitemOpen
  \bibfield  {author} {\bibinfo {author} {\bibfnamefont {F.}~\bibnamefont {Murabayashi}}, \bibinfo {author} {\bibfnamefont {R.}~\bibnamefont {Sano}}, \bibinfo {author} {\bibfnamefont {F.}~\bibnamefont {Ronetti}}, \bibinfo {author} {\bibfnamefont {J.}~\bibnamefont {Rech}}, \bibinfo {author} {\bibfnamefont {T.}~\bibnamefont {Martin}}, \bibinfo {author} {\bibfnamefont {T.}~\bibnamefont {Jonckheere}},\ and\ \bibinfo {author} {\bibfnamefont {T.}~\bibnamefont {Kato}},\ }\bibfield  {title} {\bibinfo {title} {Microwave response of fractional quantum hall droplets with quasiparticle tunneling},\ }\href {https://journals.aps.org/prb/abstract/10.1103/rngt-9l21} {\bibfield  {journal} {\bibinfo  {journal} {Physical Review B}\ }\textbf {\bibinfo {volume} {114}},\ \bibinfo {pages} {115402} (\bibinfo {year} {2026})}\BibitemShut {NoStop}%
\bibitem [{\citenamefont {Strzalkowski}\ \emph {et~al.}(1976)\citenamefont {Strzalkowski}, \citenamefont {Joshi},\ and\ \citenamefont {Crowell}}]{strzalkowski1976}%
  \BibitemOpen
  \bibfield  {author} {\bibinfo {author} {\bibfnamefont {I.}~\bibnamefont {Strzalkowski}}, \bibinfo {author} {\bibfnamefont {S.}~\bibnamefont {Joshi}},\ and\ \bibinfo {author} {\bibfnamefont {C.}~\bibnamefont {Crowell}},\ }\bibfield  {title} {\bibinfo {title} {Dielectric constant and its temperature dependence for {G}a{A}s, {C}d{T}e and {Z}n{S}e},\ }\href {https://pubs.aip.org/aip/apl/article-abstract/28/6/350/45207/Dielectric-constant-and-its-temperature-dependence?redirectedFrom=fulltext} {\bibfield  {journal} {\bibinfo  {journal} {Applied Physics Letters}\ }\textbf {\bibinfo {volume} {28}},\ \bibinfo {pages} {350} (\bibinfo {year} {1976})}\BibitemShut {NoStop}%
\bibitem [{\citenamefont {Moore}\ and\ \citenamefont {Holm}(1996)}]{moore1996infrared}%
  \BibitemOpen
  \bibfield  {author} {\bibinfo {author} {\bibfnamefont {W.}~\bibnamefont {Moore}}\ and\ \bibinfo {author} {\bibfnamefont {R.}~\bibnamefont {Holm}},\ }\bibfield  {title} {\bibinfo {title} {Infrared dielectric constant of gallium arsenide},\ }\href {https://pubs.aip.org/aip/jap/article-abstract/80/12/6939/2671/Infrared-dielectric-constant-of-gallium-arsenide?redirectedFrom=fulltext} {\bibfield  {journal} {\bibinfo  {journal} {Journal of applied physics}\ }\textbf {\bibinfo {volume} {80}},\ \bibinfo {pages} {6939} (\bibinfo {year} {1996})}\BibitemShut {NoStop}%
\bibitem [{\citenamefont {Krupka}\ \emph {et~al.}(2008)\citenamefont {Krupka}, \citenamefont {Mouneyrac}, \citenamefont {Hartnett},\ and\ \citenamefont {Tobar}}]{krupka2008}%
  \BibitemOpen
  \bibfield  {author} {\bibinfo {author} {\bibfnamefont {J.}~\bibnamefont {Krupka}}, \bibinfo {author} {\bibfnamefont {D.}~\bibnamefont {Mouneyrac}}, \bibinfo {author} {\bibfnamefont {J.~G.}\ \bibnamefont {Hartnett}},\ and\ \bibinfo {author} {\bibfnamefont {M.~E.}\ \bibnamefont {Tobar}},\ }\bibfield  {title} {\bibinfo {title} {Use of whispering-gallery modes and quasi-{TE}$_{0np}$ modes for broadband characterization of bulk gallium arsenide and gallium phosphide samples},\ }\href {https://ieeexplore.ieee.org/document/4488211} {\bibfield  {journal} {\bibinfo  {journal} {IEEE Transactions on microwave theory and techniques}\ }\textbf {\bibinfo {volume} {56}},\ \bibinfo {pages} {1201} (\bibinfo {year} {2008})}\BibitemShut {NoStop}%
\bibitem [{\citenamefont {Johnson}\ and\ \citenamefont {Vignale}(2003)}]{johnson2003}%
  \BibitemOpen
  \bibfield  {author} {\bibinfo {author} {\bibfnamefont {M.~D.}\ \bibnamefont {Johnson}}\ and\ \bibinfo {author} {\bibfnamefont {G.}~\bibnamefont {Vignale}},\ }\bibfield  {title} {\bibinfo {title} {Dynamics of dissipative quantum {H}all edges},\ }\href {https://journals.aps.org/prb/abstract/10.1103/PhysRevB.67.205332} {\bibfield  {journal} {\bibinfo  {journal} {Physical Review B}\ }\textbf {\bibinfo {volume} {67}},\ \bibinfo {pages} {205332} (\bibinfo {year} {2003})}\BibitemShut {NoStop}%
\bibitem [{\citenamefont {Talyanskii}\ \emph {et~al.}(1994)\citenamefont {Talyanskii}, \citenamefont {Simmons}, \citenamefont {Frost}, \citenamefont {Pepper}, \citenamefont {Ritchie}, \citenamefont {Churchill},\ and\ \citenamefont {Jones}}]{talyanskii1994experimental}%
  \BibitemOpen
  \bibfield  {author} {\bibinfo {author} {\bibfnamefont {V.}~\bibnamefont {Talyanskii}}, \bibinfo {author} {\bibfnamefont {M.}~\bibnamefont {Simmons}}, \bibinfo {author} {\bibfnamefont {J.}~\bibnamefont {Frost}}, \bibinfo {author} {\bibfnamefont {M.}~\bibnamefont {Pepper}}, \bibinfo {author} {\bibfnamefont {D.}~\bibnamefont {Ritchie}}, \bibinfo {author} {\bibfnamefont {A.}~\bibnamefont {Churchill}},\ and\ \bibinfo {author} {\bibfnamefont {G.}~\bibnamefont {Jones}},\ }\bibfield  {title} {\bibinfo {title} {Experimental investigation of the damping of low-frequency edge magnetoplasmons in {G}a{A}s-{A}l$_x$ {G}a$_{1- x}$ {A}s heterostructures},\ }\href {https://journals.aps.org/prb/abstract/10.1103/PhysRevB.50.1582} {\bibfield  {journal} {\bibinfo  {journal} {Physical Review B}\ }\textbf {\bibinfo {volume} {50}},\ \bibinfo {pages} {1582} (\bibinfo {year} {1994})}\BibitemShut {NoStop}%
\bibitem [{\citenamefont {Choi}\ \emph {et~al.}(1987)\citenamefont {Choi}, \citenamefont {Tsui},\ and\ \citenamefont {Alavi}}]{choi1987experimental}%
  \BibitemOpen
  \bibfield  {author} {\bibinfo {author} {\bibfnamefont {K.~K.}\ \bibnamefont {Choi}}, \bibinfo {author} {\bibfnamefont {D.~C.}\ \bibnamefont {Tsui}},\ and\ \bibinfo {author} {\bibfnamefont {K.}~\bibnamefont {Alavi}},\ }\bibfield  {title} {\bibinfo {title} {Experimental determination of the edge depletion width of the two-dimensional electron gas in {G}a{A}s/{A}l$_{x}${G}a$_{1- x}${A}s},\ }\href {https://pubs.aip.org/aip/apl/article-abstract/50/2/110/52144/Experimental-determination-of-the-edge-depletion?redirectedFrom=fulltext} {\bibfield  {journal} {\bibinfo  {journal} {Applied physics letters}\ }\textbf {\bibinfo {volume} {50}},\ \bibinfo {pages} {110} (\bibinfo {year} {1987})}\BibitemShut {NoStop}%
\bibitem [{\citenamefont {Chklovskii}\ \emph {et~al.}(1992)\citenamefont {Chklovskii}, \citenamefont {Shklovskii},\ and\ \citenamefont {Glazman}}]{chklovskii1992electrostatics}%
  \BibitemOpen
  \bibfield  {author} {\bibinfo {author} {\bibfnamefont {D.}~\bibnamefont {Chklovskii}}, \bibinfo {author} {\bibfnamefont {B.~I.}\ \bibnamefont {Shklovskii}},\ and\ \bibinfo {author} {\bibfnamefont {L.}~\bibnamefont {Glazman}},\ }\bibfield  {title} {\bibinfo {title} {Electrostatics of edge channels},\ }\href {https://journals.aps.org/prb/abstract/10.1103/PhysRevB.46.4026} {\bibfield  {journal} {\bibinfo  {journal} {Physical Review B}\ }\textbf {\bibinfo {volume} {46}},\ \bibinfo {pages} {4026} (\bibinfo {year} {1992})}\BibitemShut {NoStop}%
\bibitem [{\citenamefont {Gelfand}\ and\ \citenamefont {Halperin}(1994)}]{gelfand1994edge}%
  \BibitemOpen
  \bibfield  {author} {\bibinfo {author} {\bibfnamefont {B.}~\bibnamefont {Gelfand}}\ and\ \bibinfo {author} {\bibfnamefont {B.}~\bibnamefont {Halperin}},\ }\bibfield  {title} {\bibinfo {title} {Edge electrostatics of a mesa-etched sample and edge-state-to-bulk scattering rate in the fractional quantum {H}all regime},\ }\href {https://journals.aps.org/prb/abstract/10.1103/PhysRevB.49.1862} {\bibfield  {journal} {\bibinfo  {journal} {Physical Review B}\ }\textbf {\bibinfo {volume} {49}},\ \bibinfo {pages} {1862} (\bibinfo {year} {1994})}\BibitemShut {NoStop}%
\bibitem [{\citenamefont {Dahl}\ \emph {et~al.}(1995)\citenamefont {Dahl}, \citenamefont {Manus}, \citenamefont {Kotthaus}, \citenamefont {Nickel},\ and\ \citenamefont {Schlapp}}]{dahl1995edge}%
  \BibitemOpen
  \bibfield  {author} {\bibinfo {author} {\bibfnamefont {C.}~\bibnamefont {Dahl}}, \bibinfo {author} {\bibfnamefont {S.}~\bibnamefont {Manus}}, \bibinfo {author} {\bibfnamefont {J.}~\bibnamefont {Kotthaus}}, \bibinfo {author} {\bibfnamefont {H.}~\bibnamefont {Nickel}},\ and\ \bibinfo {author} {\bibfnamefont {W.}~\bibnamefont {Schlapp}},\ }\bibfield  {title} {\bibinfo {title} {Edge magnetoplasmons in single two-dimensional electron disks at microwave frequencies: Determination of the lateral depletion length},\ }\href {https://pubs.aip.org/aip/apl/article-abstract/66/17/2271/521499/Edge-magnetoplasmons-in-single-two-dimensional?redirectedFrom=fulltext} {\bibfield  {journal} {\bibinfo  {journal} {Applied physics letters}\ }\textbf {\bibinfo {volume} {66}},\ \bibinfo {pages} {2271} (\bibinfo {year} {1995})}\BibitemShut {NoStop}%
\end{thebibliography}%

\appendix

\section{Geometrical model of the QPC}

\begin{figure}[h]
    \centering
    \includegraphics[width=\linewidth]{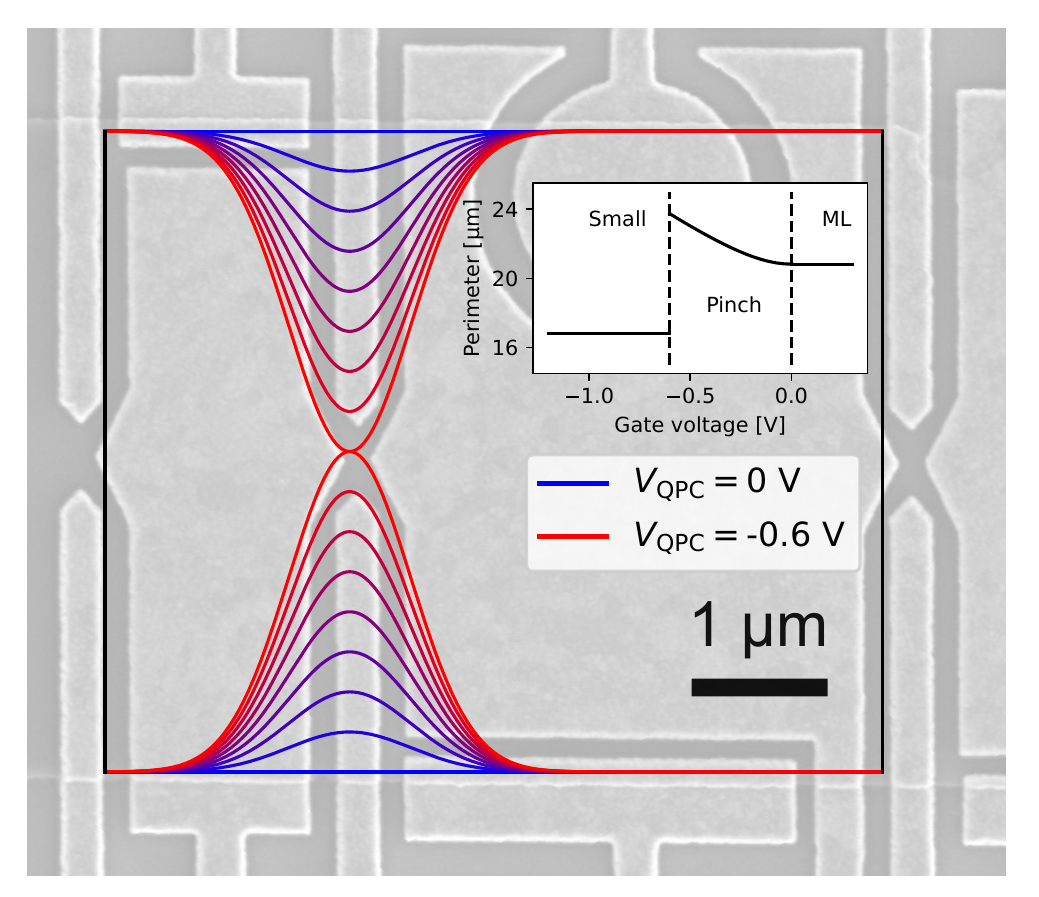}
    \caption{\textbf{Geometrical characterization of QPCs:} Assumed path of the EMPs for different values of the QPC voltage, from the open configuration (in blue) to the pinched-off configuration (in red). The inset show the corresponding perimeter of the cavity. An SEM image of the sample is placed behind the graph with the appropriate scale.}
    \label{geometrical model}
\end{figure}

In a crude approximation, our EMPs can be considered to be following equipotentials resulting from the electric field created by the QPCs. We model our system using Gaussian line shapes whose amplitude evolve linearly with the voltage in a finite range between the open configuration (\SI{0}{\volt}) and the pinched-off configuration (\SI{-0.6}{\volt}). The full length of this path is then calculated by numerically integrating these gaussian line shapes $f(x)=A\exp\left[-\frac{(x-x_0)^2}{\sigma^2} \right]$ (where $x$ is the real-space coordinate along the horizontal axis) using the following relation on the top and bottom section of the device (see figure \ref{geometrical model})
\begin{equation}
    P = \int_{x_\mathrm{\min}}^{x_\mathrm{\max}}{\sqrt{1+\left(\frac{\mathrm{d}f}{\mathrm{d}x}\right)^2}\mathrm{d}x}
\end{equation}
We use for the gaussian parameters a width $\sigma = 300$~\SI{}{\nano\meter} constant and $A = 0$ for $V_G=0$ and $A= 2.5$ for $V_G =$~\SI{-0.65}{\volt} with a linear evolution between these two points. For $V_G>0$, we assume that the EMPs follow the edges of the sample and use the geometrical parameters of our design to estimate the perimeter. Similarly, for $V_G<-0.6$\SI{}{\volt} we consider the size of the small cavity.

\section{Analytical extraction of the resonance frequency}

\begin{figure*}[t]
    \centering
    \includegraphics[width=\linewidth]{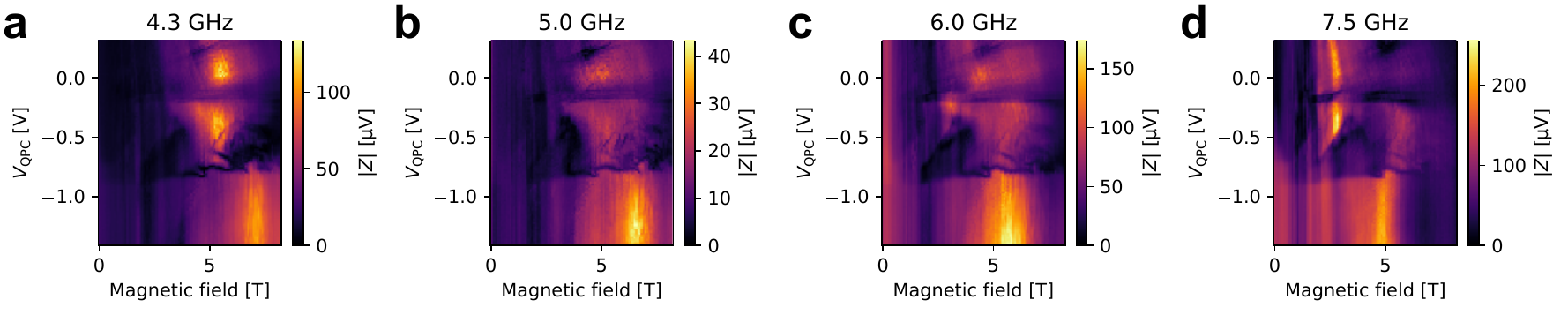}
    \caption{\textbf{Closing of QPC at constant number of electron:} \textbf{(a)-(d)} Transmission maps at fixed frequencies (\SI{4.3}{\giga\hertz}, \SI{5}{\giga\hertz}, \SI{6}{\giga\hertz}, \SI{7.5}{\giga\hertz}) as a function of the magnetic field and QPC voltage $V_\mathrm{QPC}$. These maps are taken in a configuration of constant number of electrons in the EMP cavity.}
    \label{constant_n_electron}
\end{figure*}

In order to extract a simple mathematical expression for the resonance frequency of our system, we start from the expression of the transmission coefficient obtained in \cite{frigerio2024}
\begin{equation}
    S_{ca} = \frac{t'_a t_c e^{iX_b}}{1-r'_a r_c'e^{2i X_b}}
\end{equation}
where $t'_a$, $t_c$, $r'_a$ and $r_c$ are transmission and reflection coefficients characterizing the interface between the rf gates and the 2DEG. Resonances appear in this expression when the denominator reaches local minima. Because $r'_a$ and $r'_c$ have positive real parts and negligible imaginary parts, the quantity that fixes the resonance condition is the exponential term. Resonance is then obtained when this exponential is close to $1$ and therefore
\begin{equation}
    2iX_b = 2n\pi.
\end{equation}
$X_b=k(\omega)L_b$ is defined from the EMP wave vector as well as the free propagating distance between input and output gates. In our actual device, the path between input and output gates and vice-versa are not equivalent as the gates are not centered within the cavity in the "middle left" configuration. However, reversing the chirality of the system shows no difference between both configuration and we therefore assume here that $L_b$ is average distance between both gates and we can write $2L_b = P-L_a-L_c$ where $P$ is the full perimeter of the cavity and $L_a$ and $L_c$ the width of the input and output gates respectively.

The resonance condition can then be rewritten as
\begin{equation}
    k_0 = \frac{2n\pi}{2L_b} = \frac{2n\pi}{P-L_a-L_c}.
\end{equation}
The resonance frequency in itself is obtained as
\begin{equation}
    f_n = 2n\pi\omega_0 = n\times\frac{v\mathrm{Re}(k_0)}{2\pi} = n\times\frac{v}{P-L_a-L_c}.
\end{equation}
We can now simplify the expression given for the EMP velocity by Johnson \& Vignale \cite{johnson2003} for a gated 2DEG. We then write
\begin{equation}
    v = \frac{en_Bd}{\varepsilon aB},
\end{equation}
from which we directly obtain the analytical expression of the resonance
\begin{equation}
    f_n = n\times\frac{en_Bd}{\varepsilon aB(P-L_a-L_c)}
\end{equation}

\section{Measurement methods}

As we discussed in the main text, all the data shown previously was obtained in a constant density mode. This means that when polarizing QPC$_2$ in the middle-left configuration, between each measurement point, the defining QPCs (QPC$_1$ and QPC$_3$) are open to allow for electrons to leave the cavity due to the decrease of the available area below QPC$_2$. Effectively we thus attempt to work as much as possible in a way where the electronic density is constant. However, we can bypass the opening of the outside QPCs and thus work in a configuration where electrons are trapped within the cavity for which the electronic density increases as $V_\mathrm{QPC}$ becomes more negative.

The resulting transmission maps are shown in figure \ref{constant_n_electron} for the same fixed frequencies as in the main text: \SI{4.3}{\giga\hertz}, \SI{5.0}{\giga\hertz}, \SI{6.0}{\giga\hertz} and \SI{7.5}{\giga\hertz}. Here, the frequency shift of the resonance is lest pronounced than in the constant density mode as the increase of the electronic density $n_B$ compensates partially the extension of the path of EMPs within the cavity $L$.

The same points of interest can be observed. In particular, the pinch-off point is found around the same value of $V_\mathrm{QPC}\simeq -0.75$\SI{}{\volt} and the discontinuity around $V_\mathrm{QPC}\simeq -0.1$\SI{}{\volt} is still present. Additional structure can be observed close to the pinch-off points.

However, after the pinch-off point, the signal associated to the "small" cavity remains vertical with no kink down to \SI{-1.5}{\volt}, unlike the constant density case. As we do not expect the edge states structure to evolve differently between the two configurations, we interpret this difference as the fact that in this constant electron number case, the electronic density does not vary after the pinch-off point. Therefore, this indicates that the shift observed in the "constant density" case is linked to the dynamics of electrons leaving or entering the cavity when opening QPC$_1$ and QPC$_2$.

\section{Simulated maps}

\begin{figure}[h]
    \centering
    \includegraphics[width=\linewidth]{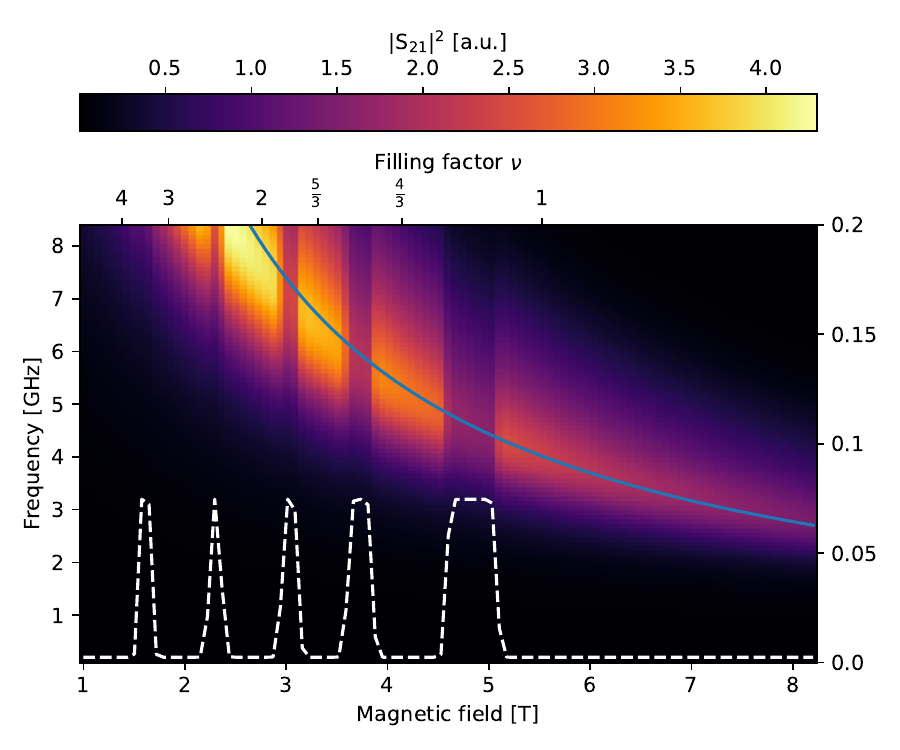}
    \caption{\textbf{Simulated rf transmission map:} Transmission of the Hall droplet as a function of frequency and magnetic field for $n = 1.31\times10^{11}$\SI{}{\per\centi\meter\squared}. The blue line is the analytical resonance expected in the limit discussed in the text from where $f=\frac{v}{P-L_a-L_c}$. The right axis refers to the dissipation term $\omega_c\tau$ which is different from zero only where the system is between Hall plateaus. In this specific configuration, we have included the $4/3$ and $5/3$ plateaus that we also observe in the experiment.}
    \label{simulated_map}
\end{figure}

On figure \ref{simulated_map}, we present the expected transmission map for our cavity in the "middle left" configuration as a function of magnetic field and frequency. The discontinuities of the map correspond to the transition between quantum Hall plateaus where we have included the presence of the fractional $\nu = 4/3$ and $5/3$ plateaus that we observe experimentally, both in the dc and rf data. The corresponding filling factors are indicated on the top axis of the figure. The dissipation term $\omega_c\tau$ included in the full expression of $k(\omega)$ \cite{frigerio2024} is plotted in white dashed line and refer to the right axis of the figure.

As we discussed in the main text of this paper, an approximation of the resonance frequency can be obtained from the resonance condition $2iX_b=2n\pi $ which leads to
\begin{equation}
    f_0 = \frac{en_Bd}{\varepsilon a(P-L_a-L_c)B}.
\end{equation}
This expression is represented on figure \ref{simulated_map} as the blue line over the 2D map. We do observe a very good agreement between the full simulated map and this analytical curve, in particular towards larger fields.

\section{Gate cross talk}

\begin{figure}[h]
    \centering
    \includegraphics[width=\linewidth]{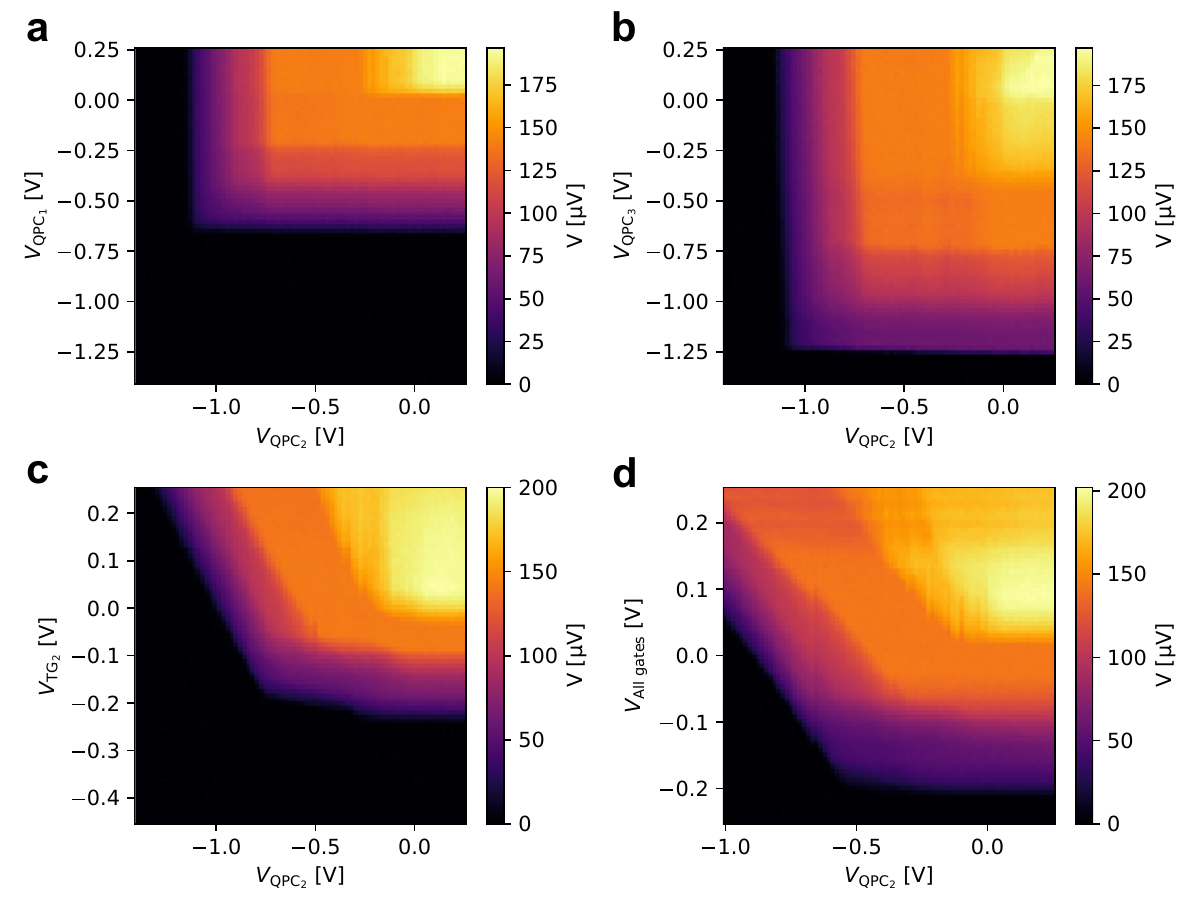}
    \caption{\textbf{DC cross talk of electrostatic gates at $\nu = 3$:} Effect of QPC1  \textbf{(a)}, QPC3  \textbf{(b)}, TG2  \textbf{(c)} and all gates  \textbf{(d)} onto the dc transmission of the device as a function of the voltage applied on QPC$_2$. These maps indicate no cross talk between QPCs but a strong effect of the top gates onto its pinch-off point.}
    \label{gate_cross_talk}
\end{figure}

In order to control the cross-talk between the many gates used in our experiment, we present in figure \ref{gate_cross_talk} various inter-gate dependences. In particular, figures \ref{gate_cross_talk}.a and b indicate that the QPCs 1 to 3 act independently of each other as seen from the absence of diagonal feature in these maps.

However, as expected, the top gates strongly impact the pinch-off point of the QPCs as can be seen for example in the case of QPC$_2$ on figures \ref{gate_cross_talk}.c and d.

\section{Fitting of 2D maps}

In order to extract the electronic density as a function of the QPC voltage we first identify the bright spots in the rf transmission map in order to locate the quantum Hall plateaus (see figure \ref{fit_examples}). This step allows us to reliably identify several plateaus both integer ($\nu = 1$ and $2$) and fractional ($\nu = 4/3$ and $2/3$). We then assume a $1/B$ dependence of the resonance frequency to get a best fit of the data and obtain the value of $f_0B$ in a given electrostatic configuration. We then use this information to extract the perimeter of the cavity shown in the main text on figure \ref{fig4_value_extraction}.

\begin{figure}
    \centering
    \includegraphics[width=0.9\linewidth]{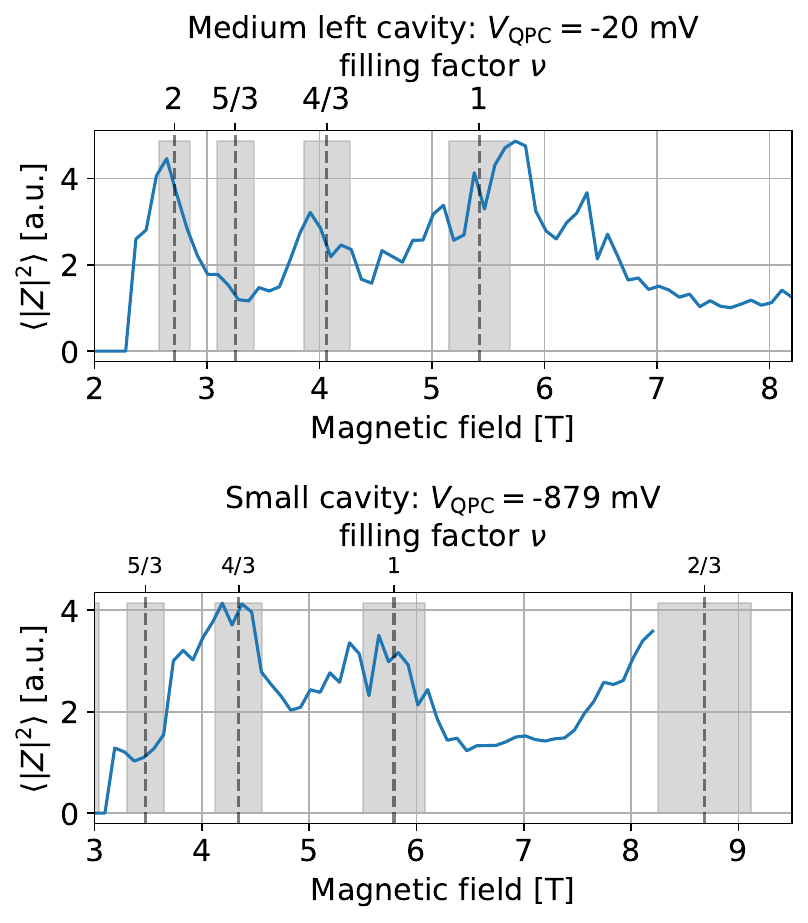}
    \caption{\textbf{Enhanced resonances on quantum Hall plateaus:} Cuts of the data shown on figure 2 of the main text, following the fitted resonance. \textbf{(a)} "Middle left" cavity \textbf{(b)} "Small" cavity. The corresponding filling factors and their associated error (grey areas) are shown on the top axis. We clearly observe peaks associated to $\nu = 1$, $\nu = 2$ and $\nu = 4/3$ on these line cuts as well as the onset of the $\nu = 2/3$ mode outside which is centered outside of our measurement window. The shift in frequency associated to the change in the cavity perimeter is also clear here.}
    \label{cuts along resonance}
\end{figure}

In figure \ref{cuts along resonance}, we present two examples of the evolution of the transmission amplitude along the resonance line with an integration bandwidth of $\Delta f= 0.075\times f$ which corresponds to the $7.5~\%$ error estimated on the determination of the constant $f_0B$. This data is extracted from figure 2 of the main text for the "middle left" and "small" configurations of the cavity. On these traces, the $\nu = 1$ and $\nu = 4/3$ quantum Hall states are clearly visible. As we mention in the main text, we also observe the tail of the resonance associated to the $\nu = 2/3$ mode that is centered outside of our accessible magnetic field range.

\section{Tuning of the antennas}

\begin{figure}[h]
    \centering
    \includegraphics[width=0.8\linewidth]{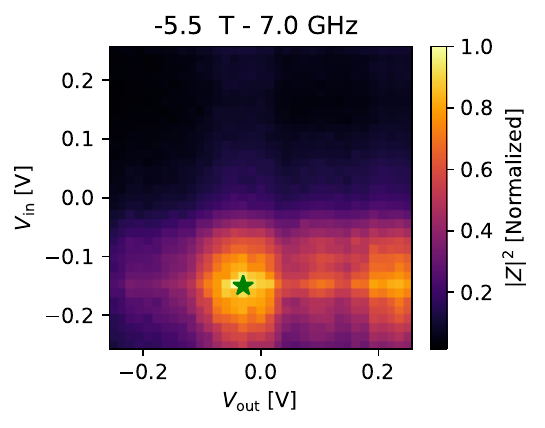}
    \caption{\textbf{Tuning of the rf antennas:} Transmission map acquired at \SI{7}{\giga\hertz} and \SI{-5.5}{\tesla} as a function of the dc voltage applied on the input and output gates. The data shown throughout this paper, was acquired in the configuration indicated by the green star at $V_\mathrm{in} = -0.15$~\SI{}{\volt} and $V_\mathrm{out}= -0.03$~\SI{}{\volt}.}
    \label{antenna tuning}
\end{figure}

In order to optimize our capacitive coupling between the edge mode and the input and output gates we plot the transmission signal as a function of the dc voltage applied on the input and output gates. This result is shown on figure \ref{antenna tuning} at $f=$~\SI{7}{\giga\hertz} and $B=$~\SI{-5.5}{\tesla}. We clearly see appear a point of maximized signal at $V_\mathrm{in} = -0.15$~\SI{}{\volt} and $V_\mathrm{out}= -0.03$~\SI{}{\volt}. We use this as an operating point throughout this paper. We observe on this map that the detection of EMP is much more sensitive to the voltage applied on the input gate than on the output one, probably due to the smaller overlap between the input gate and the 2DEG compared to that of the output gate.

\section{QPC dependence on another sample}

\begin{figure}[h!]
    \centering
    \includegraphics[width=\linewidth]{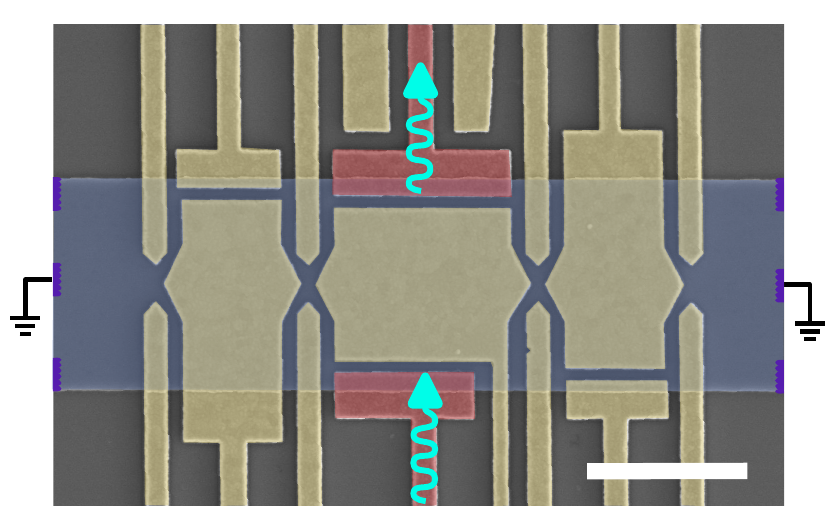}
    \caption{\textbf{Kenobi 10 - second geometry:} Colored SEM image of a Kenobi 10 sample from which data shown on figure \ref{kenobi10 exp results} were acquired. The white scale bar indicates a length of \SI{2}{\micro\meter}. The acquisition was performed on this device in the same way as on the sample shown in the main text}
    \label{kenobi10}
\end{figure}

We have reproduced our results shown in the main text of this paper on two other devices that we will label here Kenobi 10 and Kenobi 8. Kenobi 8 is a different sample from the one discussed in the main text but with the same design. Kenobi 10 has a different design shown on figure \ref{kenobi10}. This different design results in a smaller device for which the respective medium left and small cavities have perimeters $P = $~\SI{15.4}{\micro\meter} and \SI{11.4}{\micro\meter} respectively.

\begin{figure*}[h!]
    \centering
    \includegraphics[width=\linewidth]{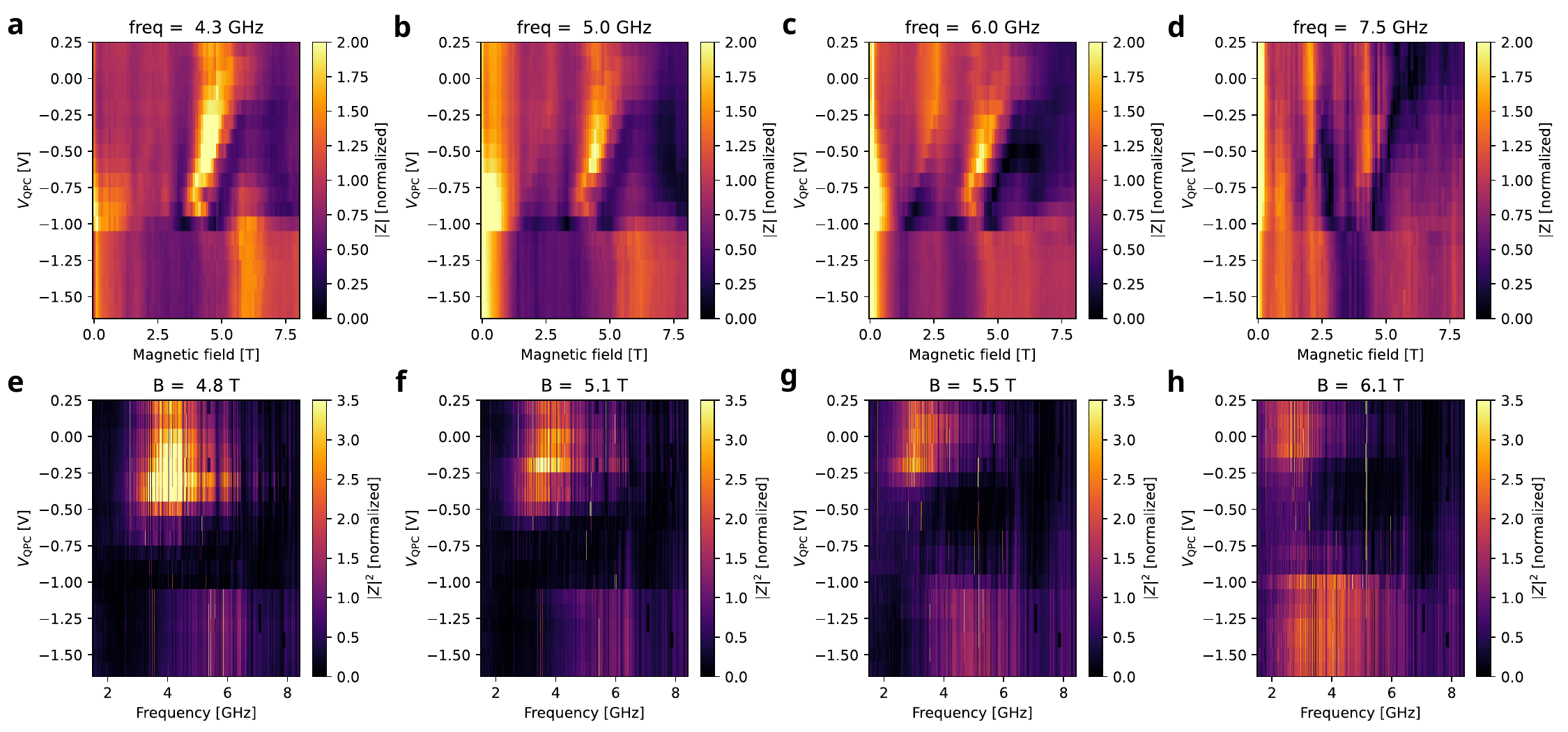}
    \caption{\textbf{Closing of QPC in sample Kenobi 10:} \textbf{(a)}-(\textbf{d}) Transmission maps from Kenobi 10 at fixed frequencies \SI{4.3}{\giga\hertz}, \SI{4.3}{\giga\hertz}, \SI{4.3}{\giga\hertz} and \SI{7.5}{\giga\hertz} as a function of the magnetic field and the voltage applied on QPC$_2$. \textbf{(e)}-(\textbf{h}) Transmission maps from Kenobi 10 at fixed magnetic fields \SI{4.8}{\tesla}, \SI{5.1}{\tesla}, \SI{5.5}{\tesla} and \SI{6.1}{\tesla} as a function of the frequency and the voltage applied on QPC$_2$. All these maps were extracted from one single sweep of all parameters and normalized according to the procedure described in \cite{frigerio2024}.}
    \label{kenobi10 exp results}
\end{figure*}

\begin{figure}[h!]
    \centering
    \includegraphics[width=\linewidth]{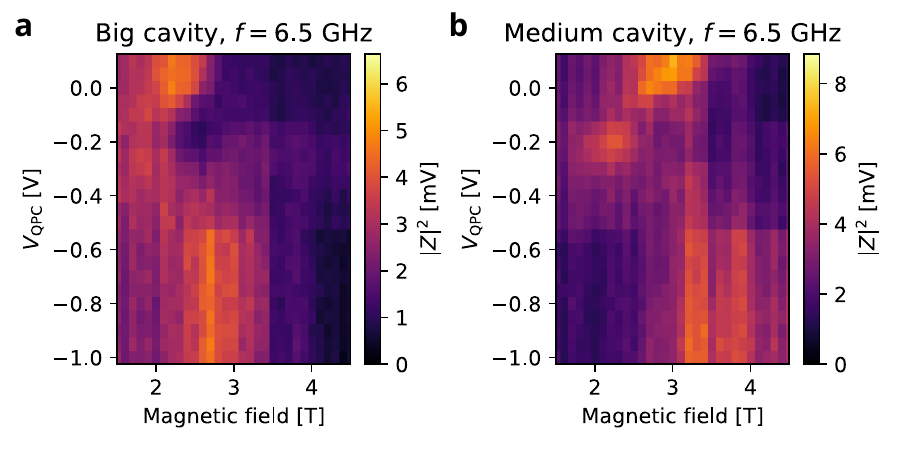}
    \caption{\textbf{Closing of QPC in sample Kenobi 8:} \textbf{(a)} Transmission map on sample Kenobi 8 in the "Big cavity" configuration at frequency \SI{6.5}{\giga\hertz} as a function of the magnetic field and the voltage applied on QPC$_2$. \textbf{(b)} Same as a. but in the medium cavity configuration. These two maps were both acquired closed to the pinch-off point that is located around $V_\mathrm{QPC}\simeq -0.4$~\SI{}{\volt}.}
    \label{kenobi8 exp results}
\end{figure}

On figures \ref{kenobi10 exp results} we present the results obtained on samples Kenobi 10 in the constant number of electron regime. The same behavior is observed here as on figure \ref{constant_n_electron}. These maps were acquired for a top gate voltage $V_\mathrm{all~gates} = -187.5$\SI{}{\milli\volt}. We can observe on maps taken at fixed frequencies (a-d) that the signal associated to "small cavity"  can be measured even before the pinch-off point. This signals the coexistence of multiple modes within our EMP cavity, a prerequisite for the emergence of cross-cavity coherence and interferometric processes. Interestingly, the shift in the resonance after pinch-off described in the main text is not seen in the Kenobi 8 and Kenobi 10 samples.

We present on figure \ref{kenobi8 exp results}, the maps acquired on sample Kenobi 8 for the fixed frequency $f=6.5$~\SI{}{\giga\hertz} in two configurations: "Big cavity" (all QPCs except the two-external ones are closed) and "medium left cavity" (same configuration as in the main text). In the case of that specific sample, the pinch-off point of the QPC is around $V_\mathrm{QPC}\simeq -0.4$~\SI{}{\volt}. The resonance in the case of the medium cavity is pushed toward larger fields, which is coherent with a smaller cavity perimeter.

\section{System parameters}

The physical parameters used for the simulation and extraction of cavity perimeters are presented in table \ref{table_params}.

\begin{table}[h]
    \centering
    \begin{tabular}{|c|c|c|c|c|c|c|}
    \hline
         $m$ & $c_G$ & $\varepsilon_r$ & $a$ & $\xi$ & $L_a$ & $L_c$ \\ 
         \hline
         $0.067\times m_0$ & 3.01\SI{}{\nano\farad\per\meter} & 12.38 & \SI{560}{\nano\meter} & \SI{612}{\nano\meter} & \SI{2.8}{\micro\meter} & \SI{1.2}{\micro\meter}\\
    \hline
    \end{tabular}
    \caption{Parameters used for the simulation of our system.}
    \label{table_params}
\end{table}

\newpage
\begin{figure*}[t]
    \centering
    \includegraphics[width=\linewidth]{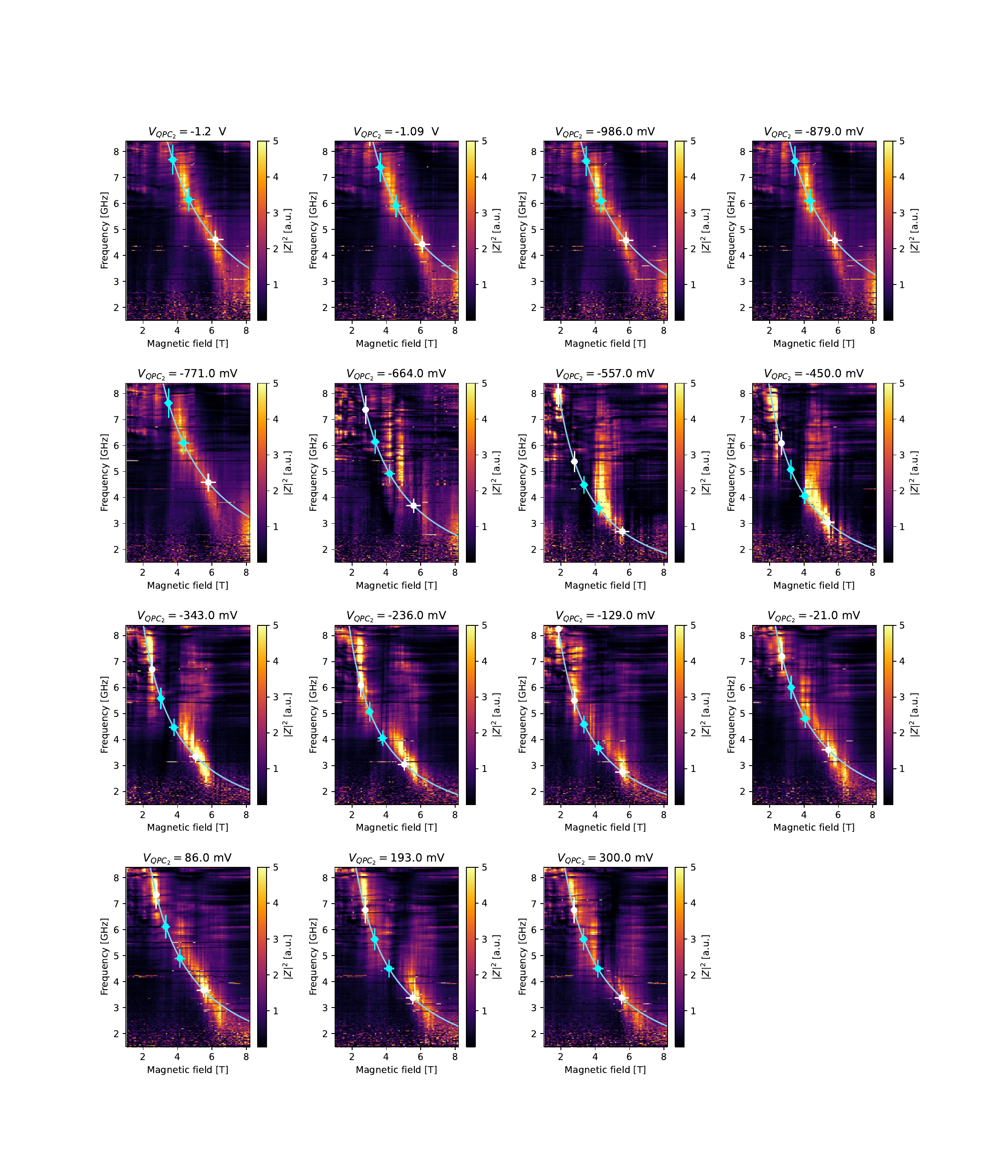}
    \caption{\textbf{Fitting procedure:} Partial sets of data used to extract the maps from figure \ref{fig4}.e to h. On top of this maps we have placed the position of quantum Hall plateaus and the $1/B$ curve from which we extract respectively the electronic density $n_B$ and the constant $f_0B$ that we plot on figure \ref{fig4_value_extraction}. Like on figure 3 of the main text, the white dots represent the integer quantum Hall plateaus while the blue dots represent the fractional ones.}
    \label{fit_examples}
\end{figure*}

\end{document}